\documentclass[10pt,a4paper,twoside]{article}
\usepackage{epsfig}
\usepackage{baltlat6}
\usepackage{dcolumn}
\usepackage{array}
\begin{document}
\ \
\vspace{0.5mm}

\setcounter{page}{111}
\vspace{8mm}

\titlehead{Baltic Astronomy, vol.\,18, 111--139, 2009}

\titleb{SPECTRAL ANALYSIS OF YSOs AND OTHER EMISSION-LINE\\ STARS IN THE
NORTH AMERICA AND PELICAN NEBULAE\\ REGION}

\begin{authorl}
\authorb{C. J. Corbally}{1},
\authorb{V. Strai\v{z}ys}{2} and
\authorb{V. Laugalys}{2}
\end{authorl}

\begin{addressl}
\addressb{1}{Vatican Observatory Research Group, Steward Observatory,
Tucson,\\ Arizona 85721, U.S.A.; corbally@as.arizona.edu}

\addressb{2}{Institute of Theoretical Physics and Astronomy, Vilnius
University,\\  Go\v{s}tauto 12, Vilnius LT-01108, Lithuania;
straizys@itpa.lt}
\end{addressl}

\submitb{Received 2009 June 20; accepted 2009 September 15}

\begin{summary} Far red spectra for 34 stars with $V$ magnitudes between
15 and 18 in the direction of the North America and Pelican nebulae
(NAP) star-forming region are obtained.  Some of these stars were known
earlier as emission-line objects, others were suspected as
pre-main-sequence stars from photometry in the {\it J, H, K}$_s$ and
{\it Vilnius} systems.  We confirm the presence of the H$\alpha$ line
emission in the spectra of 19 stars, some of them exhibit also emission
in the O\,I and Ca\,II lines.  In some of the stars the H$\alpha$
absorption line is filled with emission.  To estimate their evolutionary
status, the spectral energy distributions, based on {\it Vilnius},
2MASS, MSX and {\it Spitzer} photometry, are applied.  Only eight
emission-line stars are found to be located at a distance of the NAP
complex.  Others are either chromospherically active stars in front of
the complex or distant luminous stars with H$\alpha$ absorption and
emission components.  For five stars with faint emission the data are
not sufficient to estimate their distance.  One star is found to be a
heavily reddened K-supergiant located in the Outer arm.  The stars, for
which we failed to confirm the emission in H$\alpha$, are mostly red
dwarfs located in front of the NAP complex, two of them could be
binaries with L-type components.  Taking into account the stars
suspected to be YSOs by their 2MASS colors we conclude that the NAP
complex can possess a considerable population of young stars hidden
behind the dust cloud.  \end{summary}


\resthead{Spectral analysis of YSOs in the NAP region}
{C. J. Corbally, V. Strai\v{z}ys, V. Laugalys}

\sectionb{1}{INTRODUCTION}

The complex of two emission nebulae, known as North America and Pelican
(hereafter NAP), and the dust/molecular clouds between them, is one of
the nearest star-forming regions located at a distance of $\sim$\,500
pc.  Since Herbig's (1958) discovery it has been known to contain a
number of T Tauri and other H$\alpha$ emission stars.  The largest
number of young stars is found in the Pelican Nebula (IC 5070), a few in
the North America near Florida and in the Gulf of Mexico.  Although
later investigations in the area have revealed more emission-line stars
or young objects (Welin 1973; Gieseking 1973; Tsvetkov 1975; Gieseking
\& Schumann 1976; Marcy 1980; Kohoutek \& Wehmeyer 1997; Witham et al.
2008), the NAP complex is considered a star-forming region (SFR) not as
active as the Taurus-Auriga complex or the Orion association.  Most of
the discovered young objects are located at the edges of the dust
clouds.  The discovery by Laugalys et al.  (2006) of a group of K and M
dwarfs in the direction of the dark cloud LDN\,935, having signs of
H$\alpha$ emission, was an indication that a hidden population of young
stars could exist within or behind the dust clouds.

In the latter investigation, about 40 stars, mostly K and M
dwarfs, were suspected to be pre-main-sequence (hereafter PMS)
objects on the basis of photometric classification using seven-color CCD
photometry in the {\it Vilnius} system down to $V$ = 17--18 mag.  The
presence of H$\alpha$ emission was detected as an excessive radiation in
the $S$ filter at 656 nm, i.e., placed on the H$\alpha$ line.  For this,
the interstellar reddening-free diagrams $Q_{XZS}$ vs. $Q_{XYV}$ and
$Q_{XZS}$ vs. $Q_{YZV}$ were used (Strai\v{z}ys et al. 1998).  Many of
the suspected emission-line stars also exhibited infrared excesses, four
of them were known PMS stars.  It was supposed that the
detected stars belong to either the classical T Tauri or post-T Tauri
groups and are affected by considerable interstellar reddening.

%%%%%%%%%%%%%%%%%%%%%% TABLE 1. LIST OF OBJECTS
\begin{table}[!t]
\begin{center}
\vbox{\small\tabcolsep=4pt
\parbox[c]{124mm}{\baselineskip=10pt
{\normbf\ \ Table 1.}{\norm\ List of the investigated stars in which
emission lines or emission components were found. For the Laugalys et
al. (2006) stars, $V$ is
the green \hbox{\it Vilnius} magnitude rounded to the nearest 0.1 mag,
and $F$ is the
red photographic magnitude taken from GSC\,2.3.2 (Lasker et al. 2008).
In the last column, spectral types are from Laugalys et al.
(2006), and YSO means a suspected young stellar object selected from the
infrared sources (see the text). For them both $V$ and $F$ magnitudes
are from GSC\,2.3.2.\lstrut}}
\begin{tabular}{lccccccl}
\tablerule
 Name  & RA\,(2000) & DEC\,(2000) & $\ell$ & $b$~~~ & $V$ & $F$~~ &Type \\
\tablerule
II-64  & 20:57:07.57 & +43:41:59.7 & 84.829 & --1.170 & 17.6 & 15.3 & k-m V,T? \\  [-1pt]
II-77  & 20:57:22.25 & +43:57:53.4 & 85.059 & --1.031 & 15.1 & 13.6 & k0 V,e?  \\  [-1pt]
II-108* & 20:57:48.80 & +43:50:23.6 & 85.016 & --1.173 & 16.6 & 14.5 & k-m V,T? \\ [-1pt]
II-109 & 20:57:50.06 & +43:50:57.0 & 85.026 & --1.170 & 17.9 & 16.0 & m1 V,e?  \\  [-1pt]
II-113* & 20:57:56.51 & +43:52:36.3 & 85.059 & --1.166 & 16.3 & 14.4 & k6 V,e?  \\ [-1pt]
II-114 & 20:57:57.50 & +43:50:09.0 & 85.030 & --1.195 & 17.0 & 14.9 & m2 V,e?  \\  [-1pt]
II-117* & 20:57:59.87 & +43:53:26.0 & 85.076 & --1.165 & 15.5 & 13.3 & T?       \\ [-1pt]
II-122 & 20:58:06.05 & +43:49:33.0 & 85.039 & --1.221 & 17.7 & 15.0 & k-m V,e? \\  [-1pt]
III-85 & 20:59:35.36 & +43:52:03.7 & 85.247 & --1.397 & 16.6 & 15.1 & g:,e?    \\  [-1pt]
IV-5   & 20:53:51.88 & +44:24:10.1 & 84.986 & --0.268 & 16.7 & 15.0 & T?       \\  [-1pt]
IV-39  & 20:54:25.54 & +44:23:02.0 & 85.036 & --0.357 & 17.4 & 15.4 & k-m, e?  \\  [-1pt]
IV-62  & 20:54:45.36 & +44:33:02.8 & 85.202 & --0.295 & 17.2 & 13.4 & k0 V,e?  \\  [-1pt]
IV-98  & 20:55:11.23 & +44:24:05.3 & 85.138 & --0.450 & 16.2 & 14.2 & Be:      \\  [-1pt]
CSL\,3 & 20:48:50.70 & +43:49:49.6 & 83.971 & +0.062  &      & 16.6 & YSO \\       [-1pt]
CSL\,4 & 20:49:32.19 & +44:17:03.1 & 84.402 & +0.252  &      & 16.2 & YSO \\       [-1pt]
CSL\,5 & 20:53:38.46 & +44:28:48.4 & 85.020 & --0.188 &      & 17.0 & YSO \\       [-1pt]
CSL\,6 & 20:54:46.90 & +44:48:19.7 & 85.400 & --0.134 & 17.6 & 15.3 & YSO \\       [-1pt]
CSL\,7 & 20:56:27.82 & +45:25:23.4 & 86.062 & +0.039  & 18.4 & 16.3 & YSO \\       [-1pt]
CSL\,8 & 21:01:04.87 & +43:42:34.0 & 85.305 & --1.704 &      & 15.3 & YSO \\
\tablerule
\end{tabular}
}
\end{center}
\vskip-3mm
\parbox[c]{120mm}{\baselineskip=7pt\footnotesize
\noindent {\bf Note.}~~The three stars with the asterisked
numbers are observed and classified independently
by Laugalys et al.  (2006) in two overlapping areas,
Area II and Area III, located in the Gulf of Mexico.  Their
numbers in Table 2 and Table 3 of the 2006 paper are:  II-108 = III-1,
II-113 = III-4 and II-117 = III-6.}
\end{table}

An additional group of suspected young stellar objects (YSOs) with
infrared excesses in the NAP area has been selected using near- and
medium- infrared photometry taken from the 2MASS and MSX surveys (see
Section 2).  The IRAS survey data in most of the NAP complex are
not available, except for the part of the area with the smallest right
ascensions.  For 34 brightest stars from both groups of the suspects we
made spectral observations to confirm or disprove their possible PMS
status.  In Section 3 we describe our spectral observations and their
results.  The search for mid-infrared excesses in the spectral energy
distributions is described in Section 4 and the search for X-ray sources
in Section 5. Discussion of the results for individual stars is given in
Section 6. The last section gives a summary of the results.

%%%%%%%%%%%%%%%%%%%%%% FIGURE 1. IDENTIFICATION CHARTS

\begin{figure}[!th]
\centerline{\psfig{figure=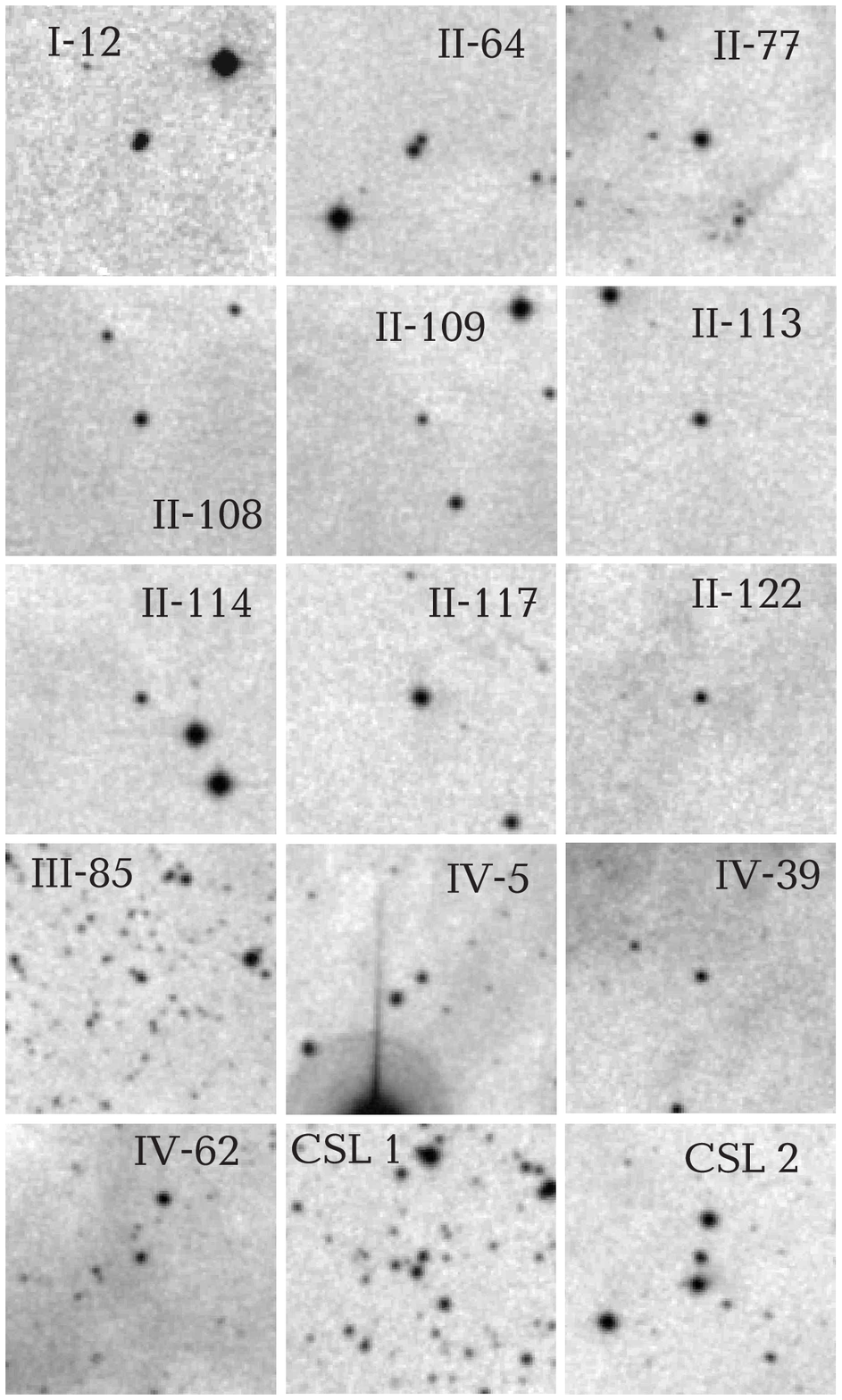,height=180mm,angle=0,clip=true}}
\vskip.5mm
\captionb{1}{Identification charts. The fields of
1.8\arcmin\,$\times$\,1.8\arcmin\ sizes are DSS2 red copies taken from
the Internet's Virtual Telescope SkyView. The identified star is always
in the center.}
\end{figure}

\begin{figure}[!th]
\centerline{\psfig{figure=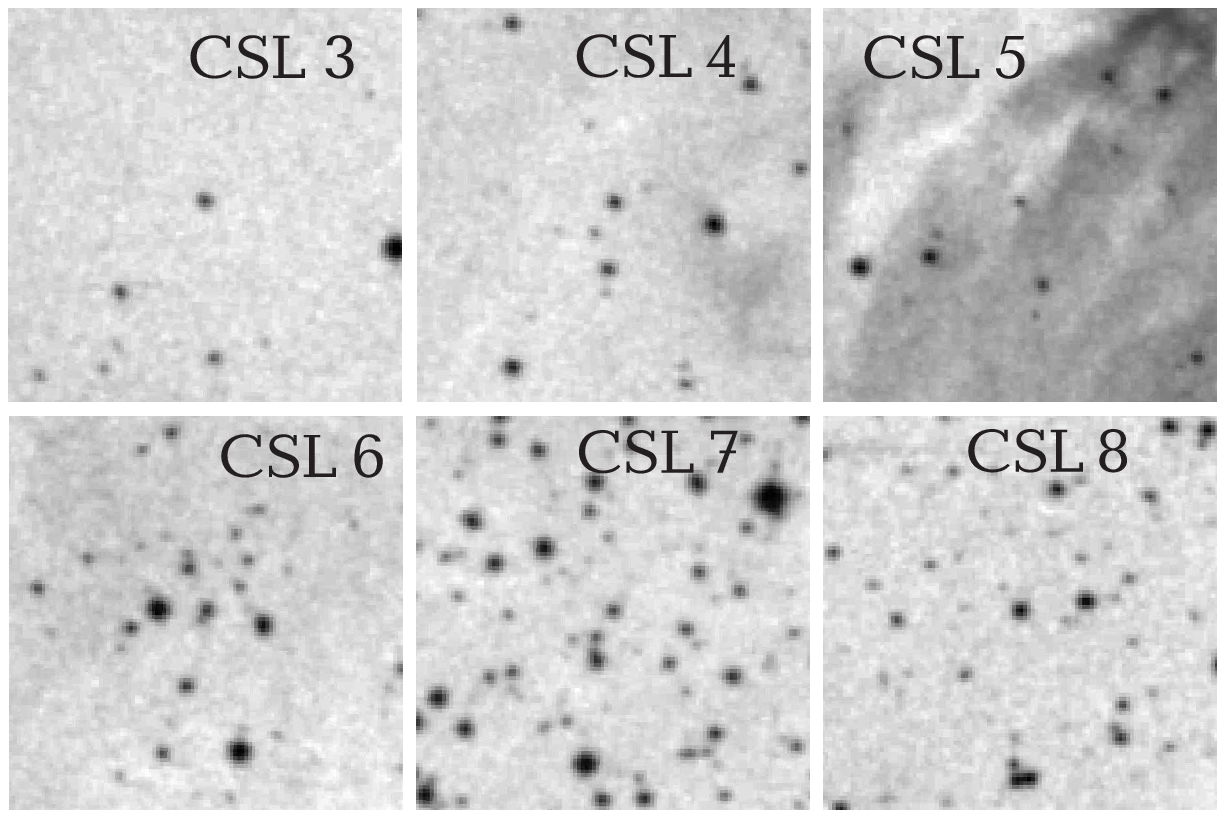,width=110mm,angle=0,clip=true}}
\vspace{0mm}
\captionc{1}{Continued.}
\end{figure}

\sectionb{2}{YSOs SUSPECTED FROM INFRARED SURVEYS}

The list of suspected YSOs was obtained in the way similar to that
applied in the Camelopardalis dark clouds (Strai\v{z}ys \& Laugalys
2007).

1. From the 2MASS catalog we selected 98\,924 stars in a
3\degr\,$\times$\,3\degr\ area with the center close to the Comer\'on \&
Pasquali (2005) O-type star (J2000.0:  20:56 = 314.0\degr, +44.0\degr)
and having errors of the $J$, $H$ and $K_s$ magnitudes $\leq$\,0.05 mag.

2. From this sample, 80 stars, satisfying the condition $Q_{JHK_s}~\leq
0.0$, were selected. Here,
\begin{equation}
Q_{JHK_s} = (J - H) - (E_{J-H}/E_{H-K_s})\,(H-K_s)~,
\end{equation}
with the ratio of color excesses $E_{J-H}/E_{H-K_s}$ = 2.0 taken
from Strai\v{z}ys et al. (2008).

Apart from YSOs, this list contains also heavily reddened O--B stars,
oxygen-rich AGB stars (SR variables, Miras and OH/IR objects),
carbon-rich AGB stars of spectral type N and cool red dwarfs of spectral
types M5--M10 and L. The problem of recognizing stars of these types was
discussed in our earlier paper devoted to search for O-type stars behind
the LDN\,935 cloud (Strai\v{z}ys \& Laugalys 2008).  Miras, carbon stars
and non-stellar sources were excluded on account of information from the
Simbad database.  Unfortunately, many objects in the list of YSO
suspects are absent in the Simbad identifier of astronomical objects.
However, their near red and far red magnitudes are available in the
IPHAS survey (Drew et al. 2005; Gonz\'alez-Solares et al. 2008).
Magnitudes or fluxes of some of the objects have been also found in the
MSX (Egan et al. 2003) and the photographic GSC\,2.3.2 catalog (Lasker
et al. 2008) at Simbad.

From the list of YSOs, suspected by their 2MASS colors, we took
only eight brightest objects down to magnitude $r$ = 17.  These stars
are denoted by initials of the authors (CSL) and the number
(from 1 to 8). The fainter objects were not practical for spectral
observations with the Steward 2.3 m telescope.

\sectionb{3}{FAR-RED SPECTRAL OBSERVATIONS}

For spectral observations we selected 34 stars.  Among these, emission
in H$\alpha$ was found in 19, including six spectra where the
absorption line is filled by emission, and in 16 stars no sign of
emission was seen.  Table 1 gives the list of stars with H$\alpha$
emission; they are identified in Figure 1. The objects for which
emission lines were absent are listed in Table 3.

Spectra for most stars were taken on 2007 October 21--23 with the Boller
\& Chivens spectrograph on the Steward Observatory 2.3 m telescope at
Kitt Peak, with a resolution of 5.7 \AA\ and a range from 607 to 939 nm
on the BCSpec 1200\,$\times$\,800 CCD detector.  The slit width was
1.5\arcsec.  Spectra of the CSL stars and of IV-62, IV-88 and IV-98 were
taken on 2008 October 19--21 with the same spectrograph and detector
setup.  Reductions of the spectra are described in our previous papers
which present the results for suspected YSOs in the Camelopardalis
clouds (Corbally \& Strai\v{z}ys 2008, 2009).

The spectra with emission signs in H$\alpha$ are shown in a widened form
in Figure 2 and their energy distributions in Figures 3 (a--r).
The equivalent widths of the prominent emission features (H$\alpha$,
Ca\,II triplet, O\,I at 8498 \AA\ and P9 at 9226 \AA), measured with the
IRAF `splot' utility, are given in Table 2. The emission in the
H$\alpha$ line in some stars does not appear above the continuum but
fills the absorption line or looks like an emission component within the
absorption line.  The last column of Table 2 gives the spectral type
obtained using the criteria described by Danks \& Dennefeld (1994).

\sectionb{4}{SPECTRAL ENERGY DISTRIBUTIONS IN THE INFRARED}

The most important confirmation that a star belongs to PMS stars (or
YSOs) is the presence of continuous emission in the infrared range with
$\lambda$\,$>$\,3 $\mu$m, which originates from dust envelopes or disks.
Therefore we constructed for the investigated stars the spectral energy
distributions (SEDs) in the form $\log \lambda F_{\lambda}$
vs.\,$\lambda$ from 0.55 to 24 $\mu$m, taking their photometry results
in different systems.  Here $\lambda$ is in $\mu$m and $F_{\lambda}$ in
erg\,$\times$\,cm$^{-2}$\,$\times$\,s$^{-1}$\,$\times$\,$\mu$m$^{-1}$.
The $V$ magnitudes were taken from {\it Vilnius} photometry (Laugalys et
al. 2006) or from GSC\,2.3.2.  The $r$ and $i$ magnitudes are from
IPHAS, the infrared magnitudes are from 2MASS, MSX and {\it Spitzer}
surveys.  The {\it Spitzer} $I_1$, $I_2$, $I_3$, $I_4$ and $M_1$ data
were taken from Guieu et al.  (2009) or were kindly communicated by L.
M. Rebull.

%%%%%%%%%%%%  FIGURE 2. WIDENED SPECTRA ALL IN ONE

\vbox{
%\begin{figure}[!th]
\vspace{3mm}
\centerline{\psfig{figure=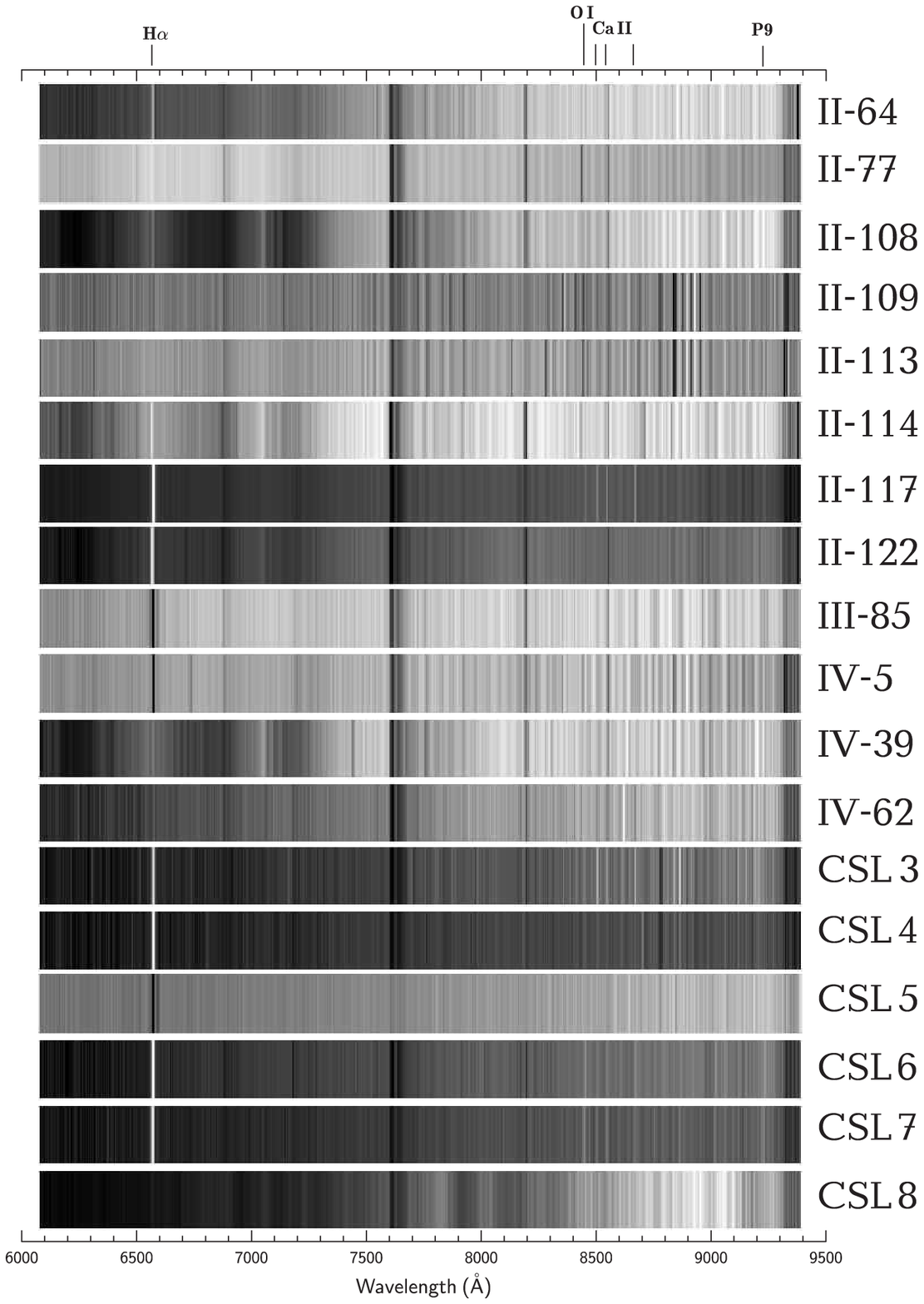,width=124mm,angle=0,clip=true}}
\vspace{0mm}
\captionb{2}{The widened spectra of stars with the emission in
H$\alpha$. The telluric H$_2$O and O$_2$ bands are not excluded.
}}
%\end{figure}
\newpage

%%%%%%%%%%%%%%%% FIGURE 3. SPECTRAL SCANS

\vbox{\begin{center}
\psfig{figure=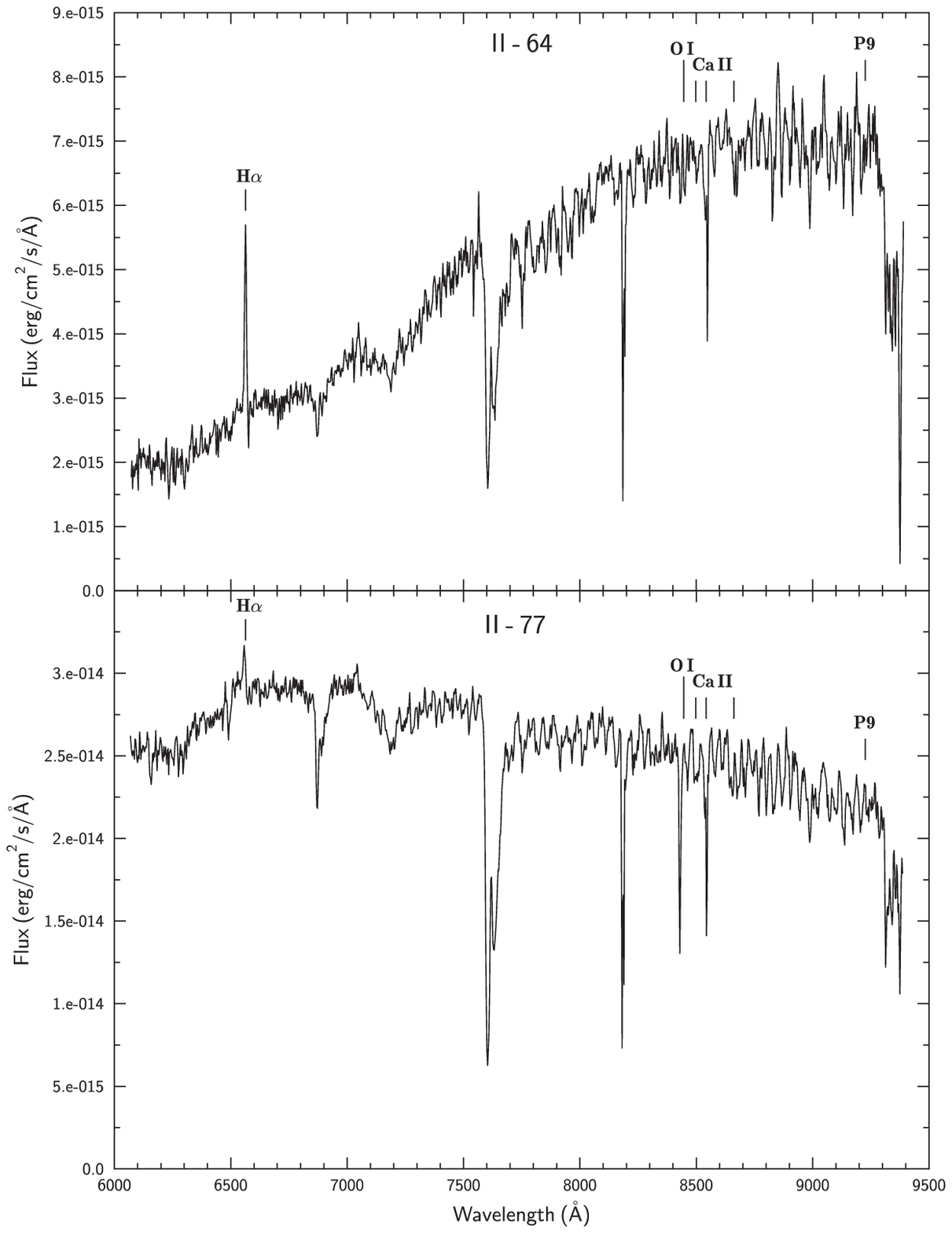,width=125mm,angle=0,clip=true}
\vskip2mm
\captionc{3a and 3b}{Spectra of the stars II-64 and II-77.}
\end{center}
}

\vbox{\begin{center}
{\psfig{figure=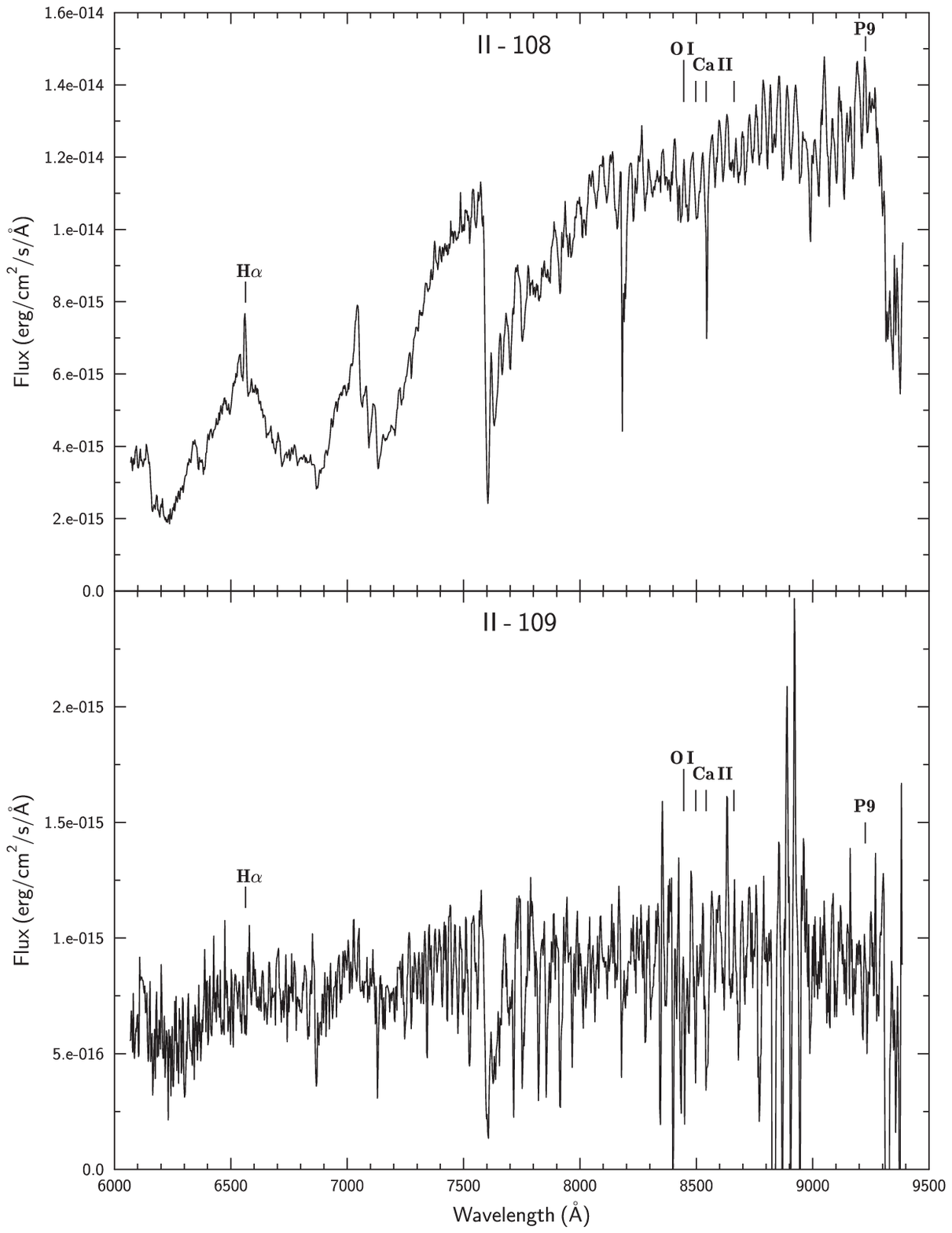,width=125mm,angle=0,clip=true}}
\vskip2mm
\captionc{3c~and~3d}{Spectra of the stars II-108 and II-109.}
\end{center}
}

\vbox{\begin{center}
{\psfig{figure=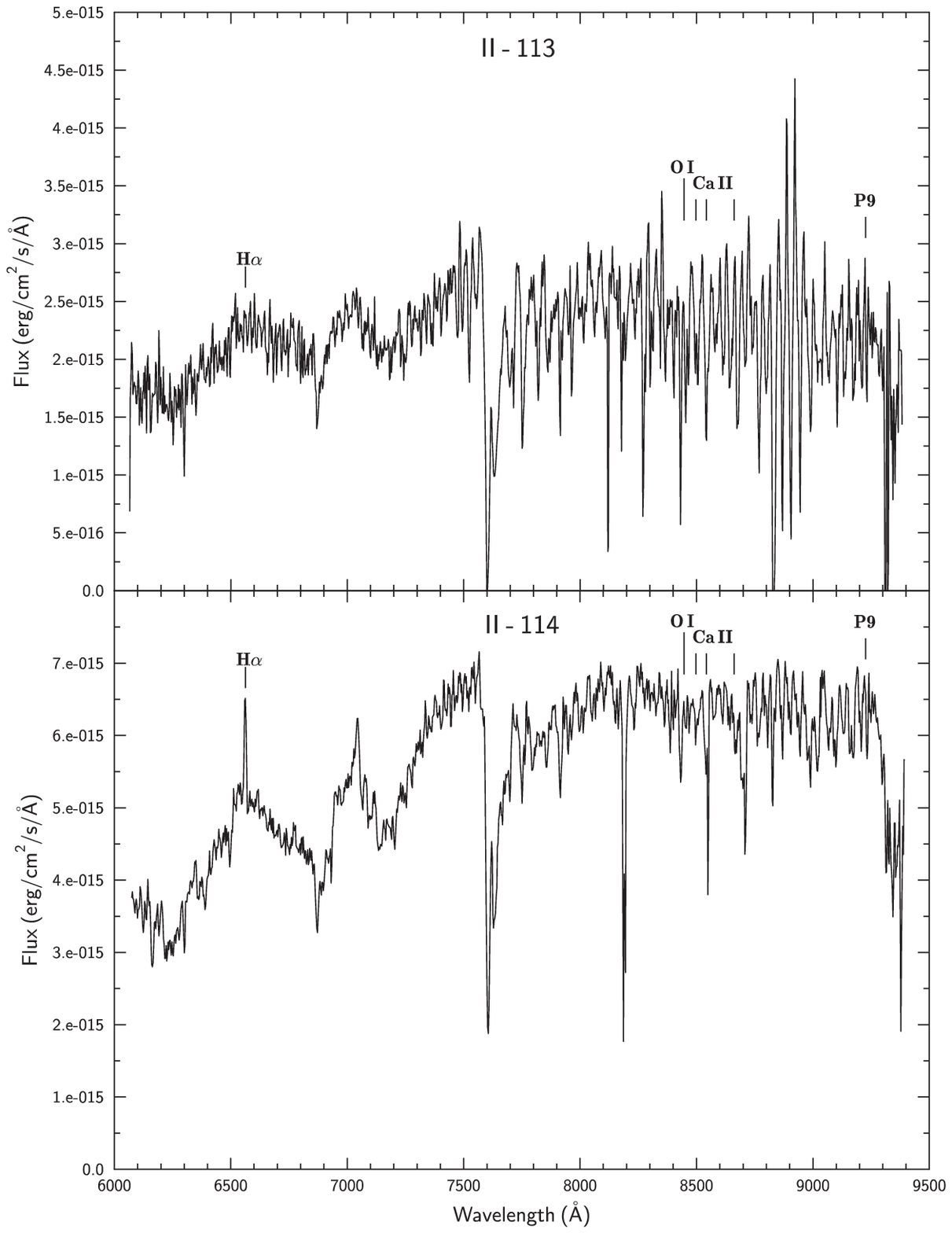,width=125mm,angle=0,clip=true}}
\vskip2mm
\captionc{3e~and~3f}{Spectra of the stars II-113 and II-114.}
\end{center}
}

\vbox{\begin{center}
{\psfig{figure=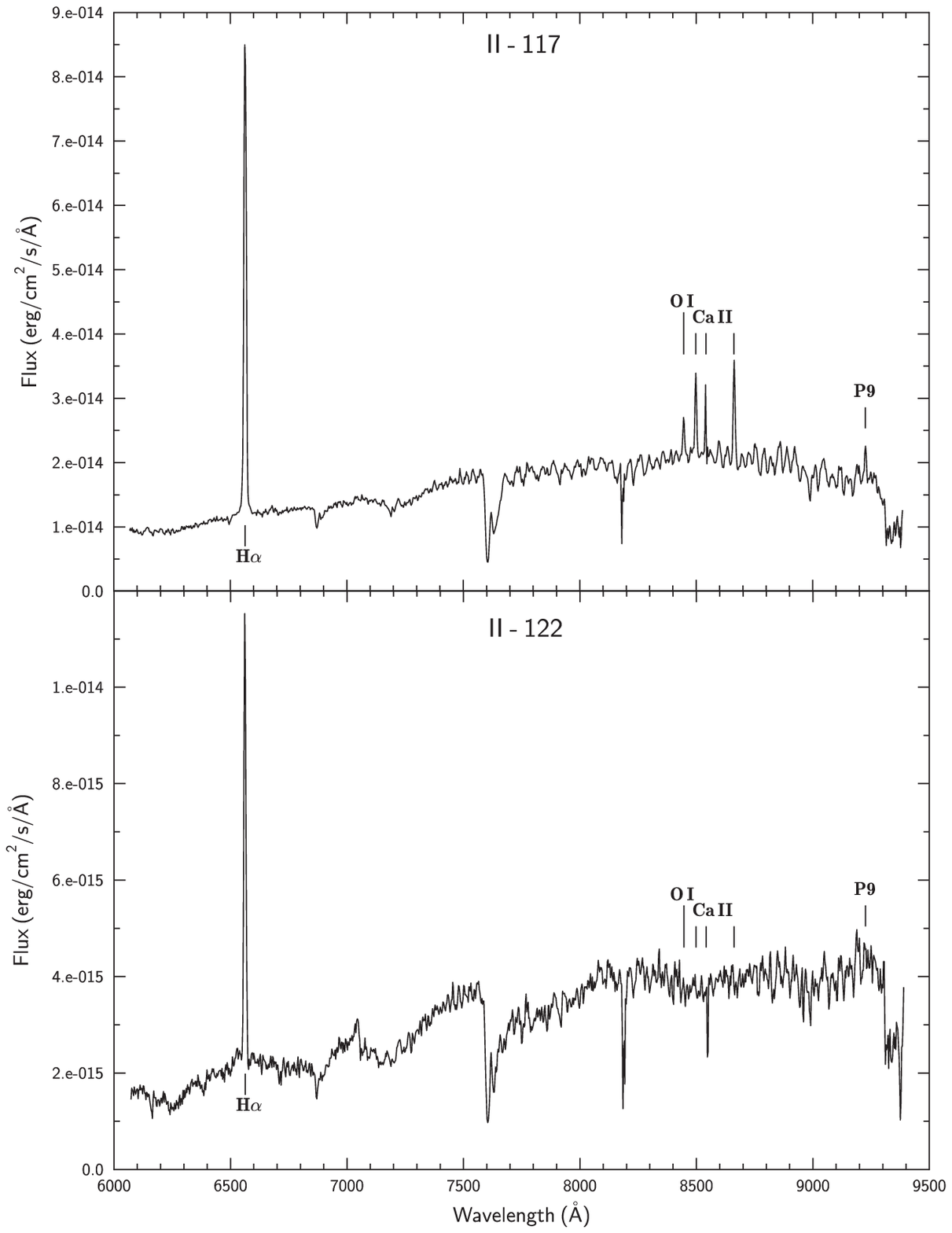,width=125mm,angle=0,clip=true}}
\vskip2mm
\captionc{3g~and~3h}{Spectra of the stars II-117 and II-122.}
\end{center}
}

\vbox{\begin{center}
{\psfig{figure=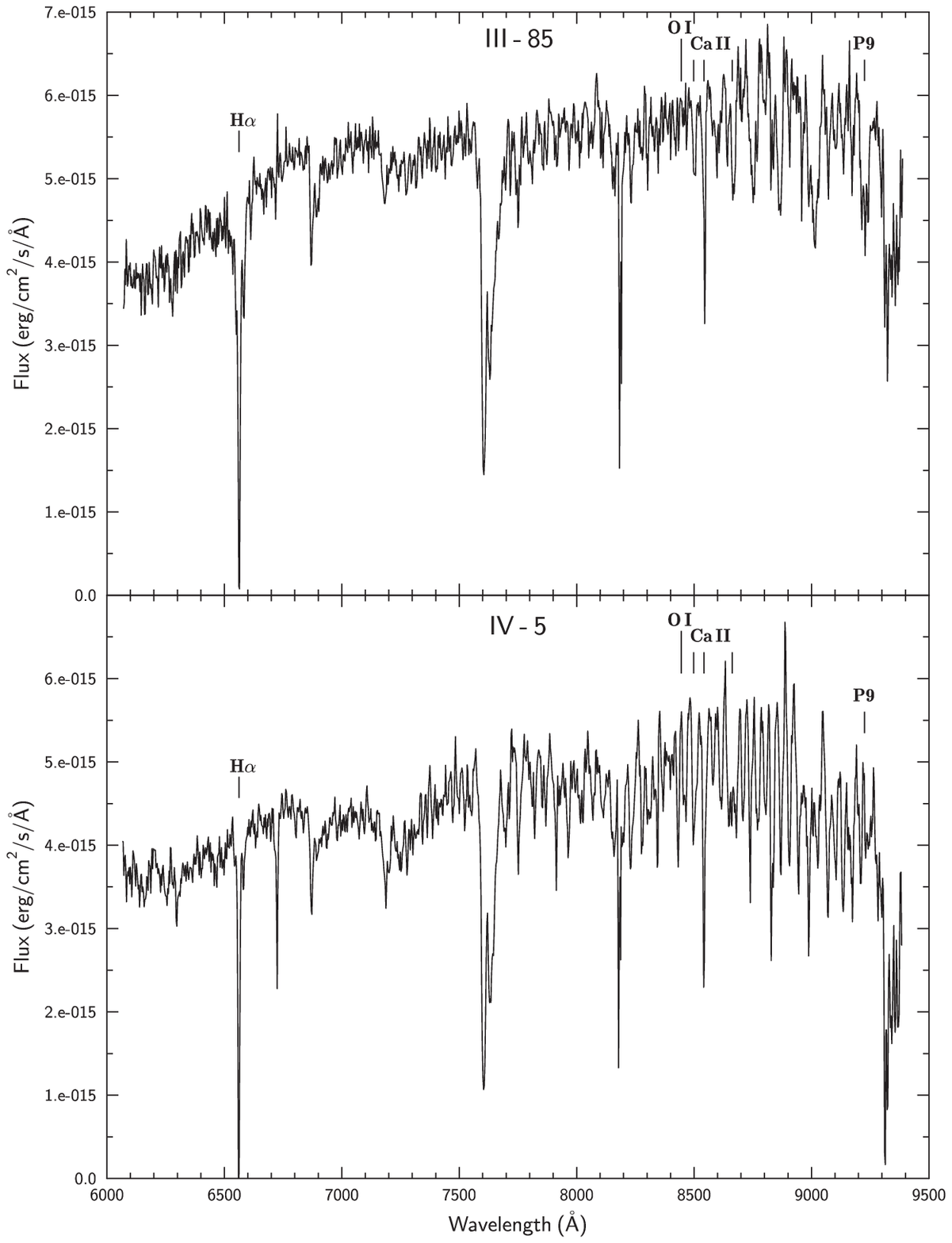,width=125mm,angle=0,clip=true}}
\vskip2mm
\captionc{3i~and~3j}{Spectra of the stars III-85 and IV-5.}
\end{center}
}

\vbox{\begin{center}
{\psfig{figure=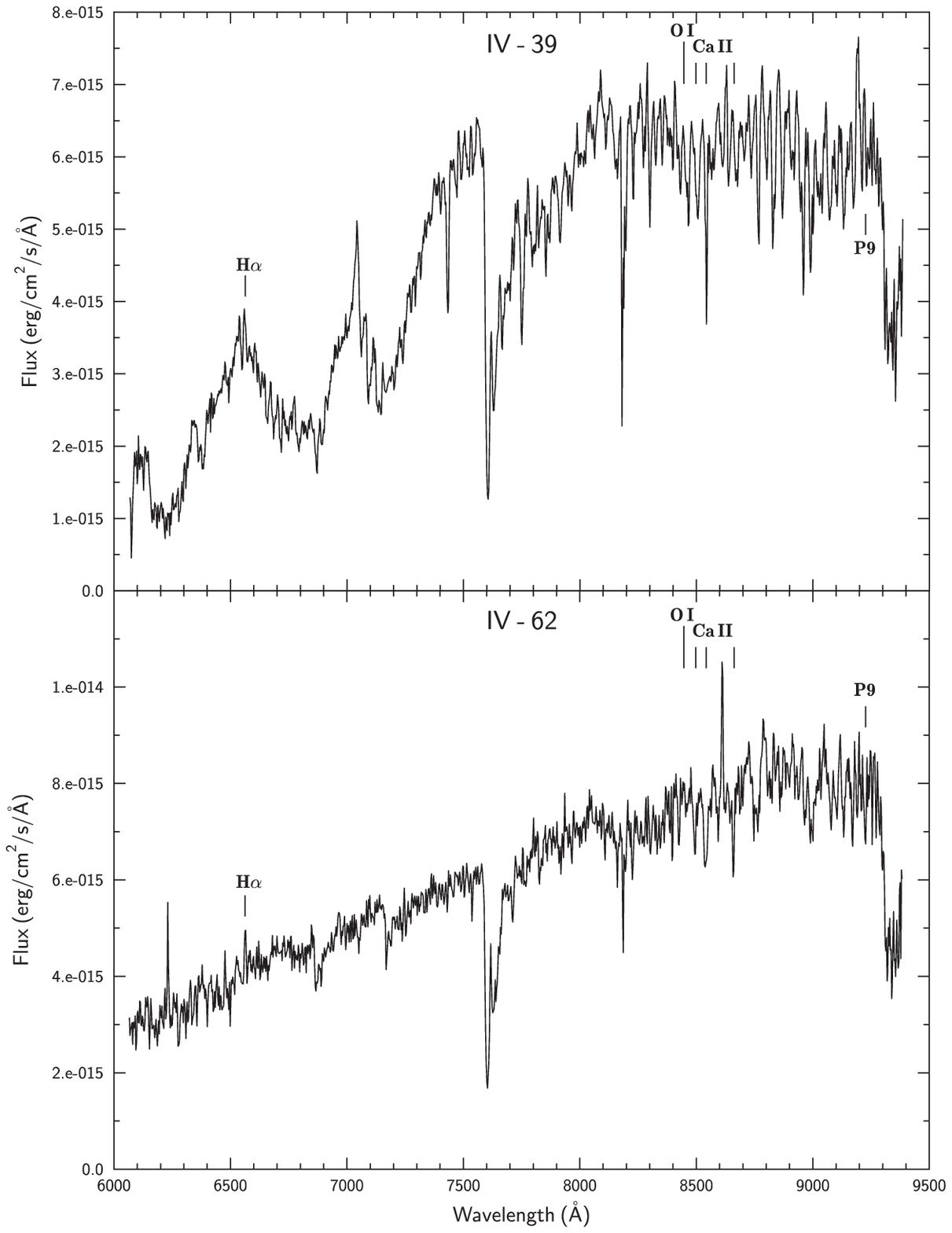,width=125mm,angle=0,clip=true}}
\vskip2mm
\captionc{3k~and~3l}{Spectra of the stars IV-39 and IV-62.}
\end{center}
}

\vbox{\begin{center}
\psfig{figure=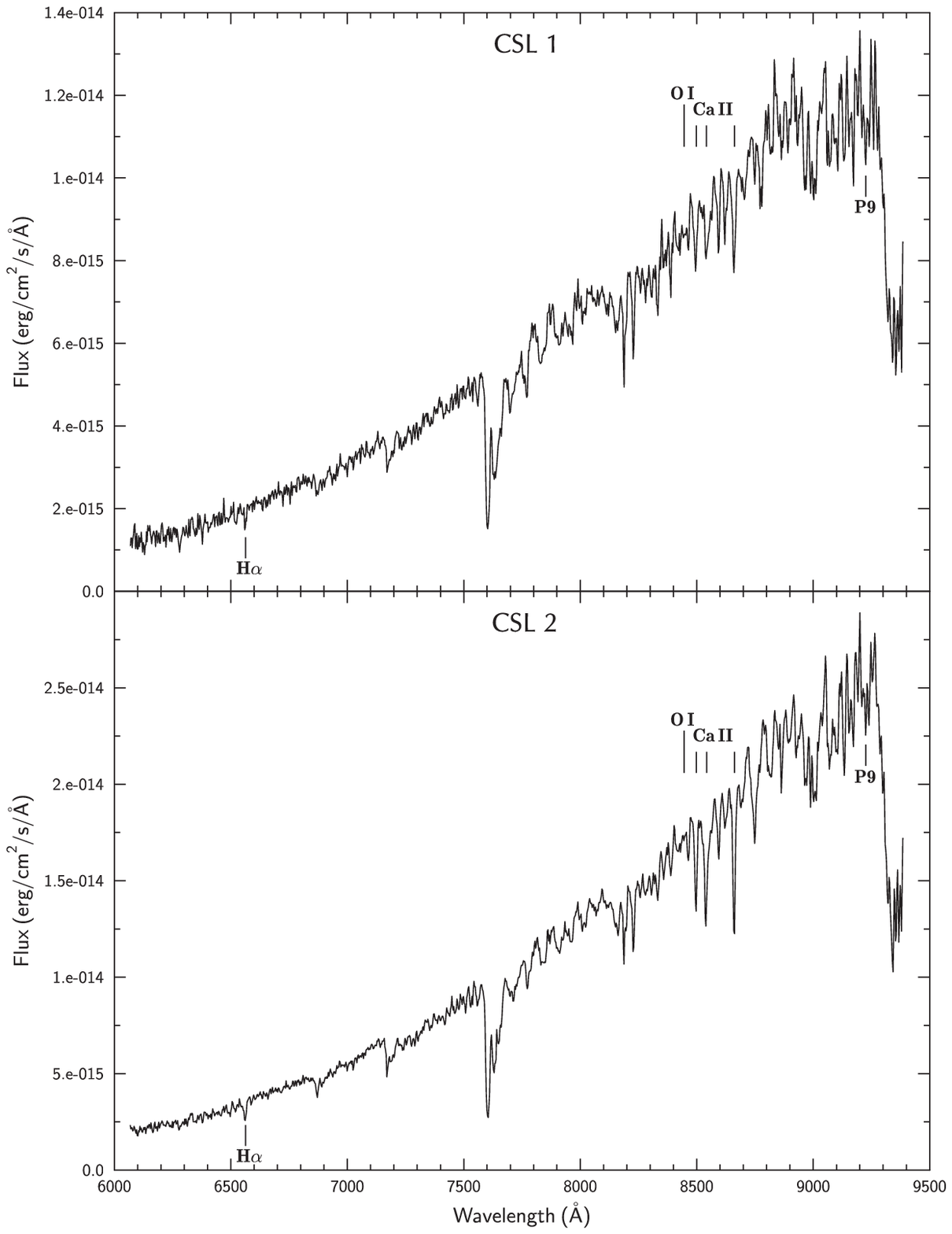,width=125mm,angle=0,clip=true}
\vskip2mm
\captionc{3m~and~3n}{Spectra of the stars CSL\,1 and CSL\,2.}
\end{center}
}
\vbox{\begin{center}
\psfig{figure=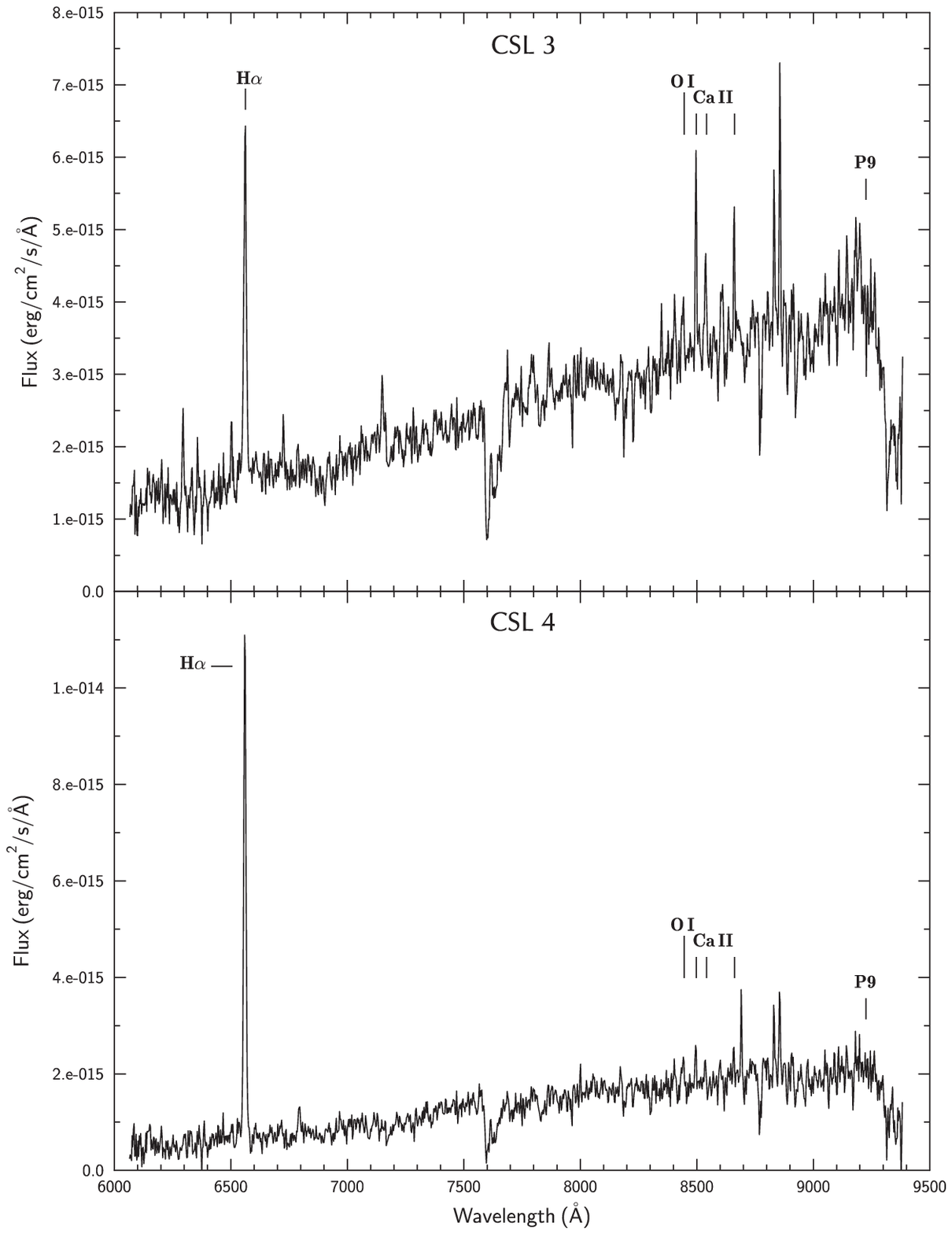,width=125mm,angle=0,clip=true}
\vskip2mm
\captionc{3o~and~3p}{Spectra of the stars CSL\,3 and CSL\,4.}
\end{center}
}

\vbox{\begin{center}
\psfig{figure=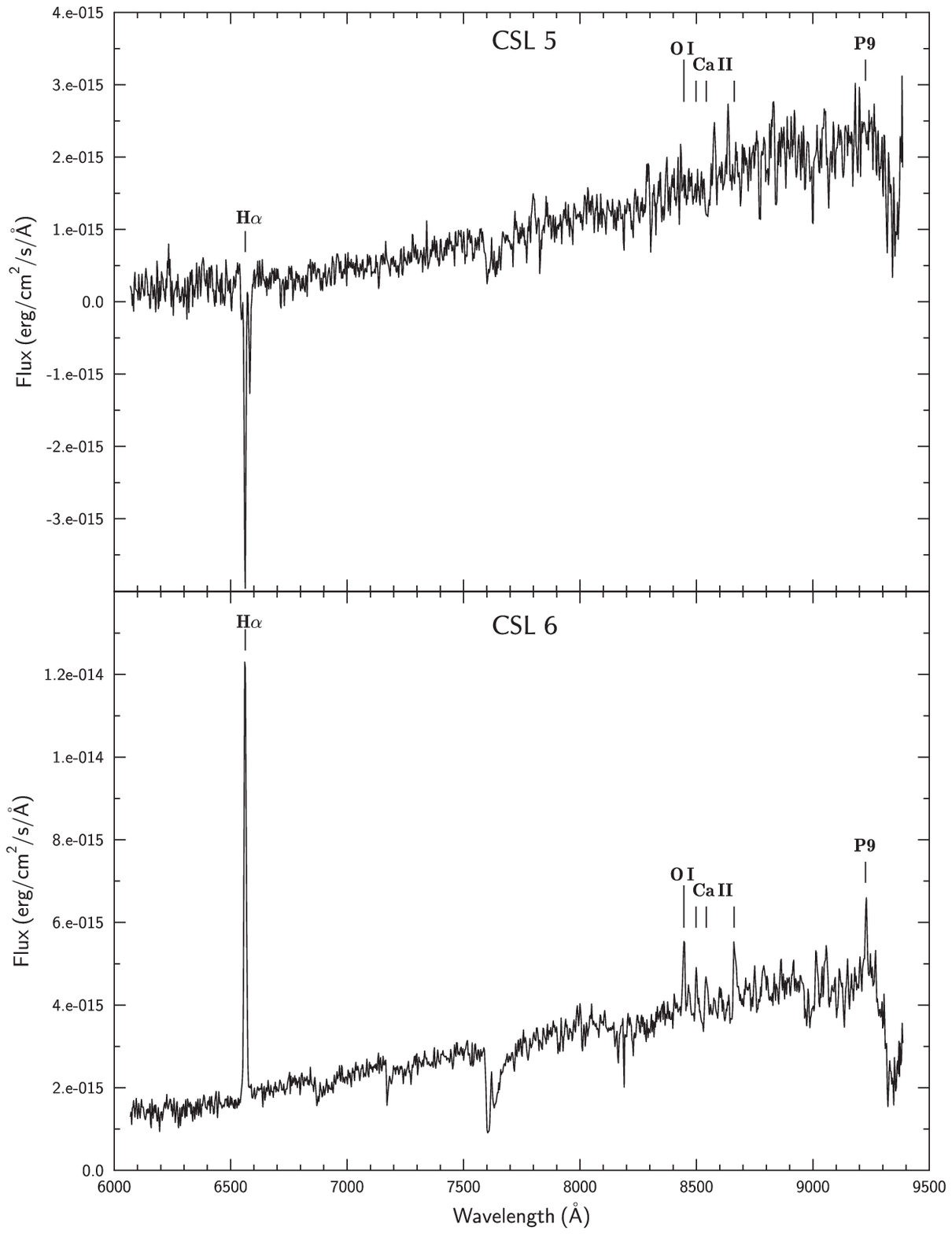,width=125mm,angle=0,clip=true}
\vskip2mm
\captionc{3q~and~3r}{Spectra of the stars CSL\,5 and CSL\,6.}
\end{center}
}

\vbox{\begin{center}
\psfig{figure=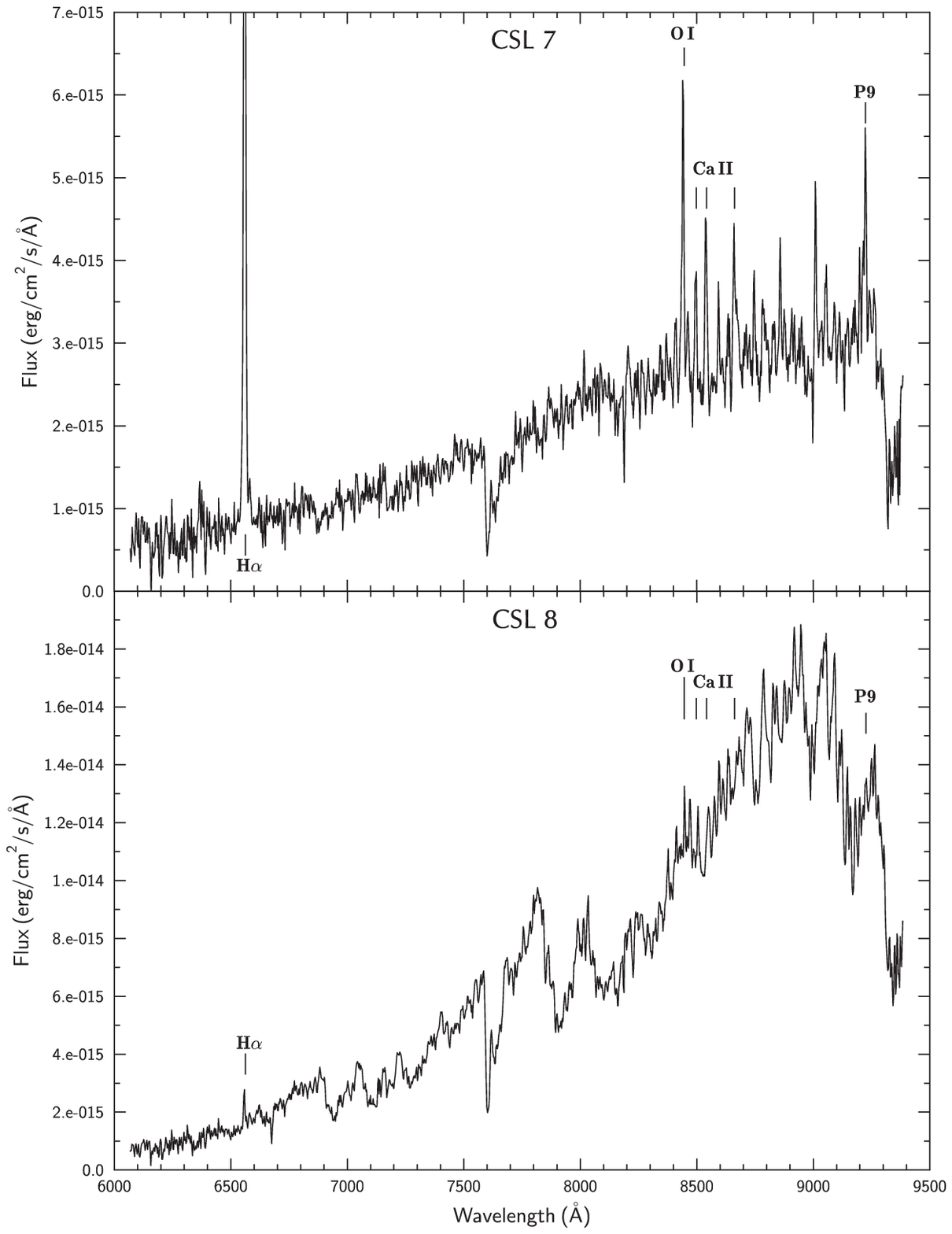,width=125mm,angle=0,clip=true}
\vskip2mm
\captionc{3s~and~3t}{Spectra of the stars CSL\,7 and CSL\,8.}
\end{center}
}

\newpage

%%%%%%%%%%%%%%%%%%%%%%  TABLE 2. EQUIVALENT WIDTHS

\begin{table}[!th]
\begin{center}
\vbox{\footnotesize\tabcolsep=2pt
\parbox[c]{110mm}{\baselineskip=10pt
{\normbf\ \ Table 2.}{\norm\ Equivalent widths of emission lines and
spectral classification.}}
\begin{tabular}[t]{lrD..{-1}D..{-1}D..{-1}D..{-1}D..{-1}D..{-1}l}
%\begin{tabular}{lrccccccl}
\tablerule
\multicolumn{1}{c}{Star}  &
\multicolumn{1}{c}{Obs. date} &
\multicolumn{1}{c}{{\it EW}\,6563} &
\multicolumn{1}{c}{{\it EW}\,8446} &
\multicolumn{1}{c}{{\it EW}\,8498} &
\multicolumn{1}{c}{{\it EW}\,8542} &
\multicolumn{1}{c}{{\it EW}\,8662}&
\multicolumn{1}{c}{{\it EW}\,9226} &
\multicolumn{1}{c}{Spectral}  \\
      &  & \multicolumn{1}{c}{H$\alpha$} &
\multicolumn{1}{c}{O\,I} &
\multicolumn{1}{c}{Ca\,II} &
\multicolumn{1}{c}{Ca\,II} &
\multicolumn{1}{c}{Ca\,II} &
\multicolumn{1}{c}{P9} &
\multicolumn{1}{c}{class} \\
\tablerule
II-64  & 2007-10-21 & -6.8  &       &       &       &       &         & K2e\,(0.5) \\[2pt]
II-77  & 2007-10-22 & -0.7  &       &       &       &       &         & G8e\,(0.1) \\[2pt]
II-108 & 2007-10-21 & -0.3  &       &       &       &       &         & M3.5e\,(0.1) \\[2pt]
II-109 & 2007-10-22 &       &       &       &       &       &         & Ge           \\[2pt]
II-113 & 2007-10-22 &       &       &       &       &       &         & Ge           \\[2pt]
II-114 & 2007-10-23 & -1.8  &       &       &       &       &         & M1e\,(0.2)   \\[2pt]
II-117 & 2007-10-21 & -79.4 & -1.8  & -4.0  & -1.4  & -5.1  & -1.2    & G5\,e\,(2) \\[2pt]
II-122 & 2007-10-23 & -41.1 &       &       &       &       &         & K7e\,(1.5)    \\[2pt]
III-85 & 2007-10-23 &       &       &       &       &       &         & A0e      \\[2pt]
IV-5   & 2007-10-21 &       &       &       &       &       &         & A0e      \\[2pt]
IV-39  & 2007-10-21 & -1.2  &       &       &       &       &         & M3e      \\[2pt]
IV-62  & 2008-10-21 & -1.8  &       &       &       &       &         & K0e      \\[2pt]
IV-98  & 2008-10-21 &       &       &       &       &       &         & B5(e)    \\[2pt]
CSL\,3 & 2008-10-19 & -30.5 &       & -4.1  & -1.9  & -2.4  &         & G0e\,(2)  \\[2pt]
CSL\,4 & 2008-10-19 & -151.3 &      & -2.9  & -1.7  & -5.4  &         & F0e\,(4)  \\[2pt]
CSL\,5 & 2008-10-21 &       &       &       &       &       &         & A5(e)     \\[2pt]
CSL\,6 & 2008-10-20 & -82.9 & -3.2  &       &       & -3.9  &         & A0e\,(3)  \\[2pt]
CSL\,7 & 2008-10-20 & -168.7 & -11.3 & -3.1 & -5.8 &        & -4.2    & A0e\,(4)  \\[2pt]
CSL\,8 & 2008-10-20 & -5.0 &        &       &       &       &         & K3\,Ie\,(3) \\[2pt]
\tablerule
\end{tabular}
}
\end{center}
\vskip-3mm
\parbox[c]{120mm}{\baselineskip=7pt\footnotesize
\noindent {\bf Notes.}~~Almost in all spectra nebular
emission is seen, it is subtracted during the
reductions; in stars II-109 and II-113 H$\alpha$ is filled in;
in stars III-85, IV-5, IV-98 and CSL\,5 the emission is seen in
H$\alpha$ core.}
\vskip-.2mm
\end{table}

The method of obtaining the SEDs from the GSC photometric data is
described by Strai\v{z}ys \& Laugalys (2007).  For the $V$ passband we
used $\lambda$ = 0.55 $\mu$m and $F_{\lambda}^{m=0}$ =
3.66\,$\times$\,10$^{-5}$.

The IPHAS $r'$ and $i'$ magnitudes (hereafter we will denote them simply
$r$ and $i$) are measured with the filters similar to the corresponding
Sloan Digital Sky Survey (SDSS) filters with central wavelengths 624 nm
and 774 nm.  However, the IPHAS magnitudes are given in the Vega-based
zero-magnitude scale while the SDSS magnitudes use the so-called AB
magnitude scale related to the energy flux units Janskys:
\begin{equation}
F_{\nu}\,({\rm Jy}) = 3631\,{\rm dex}\,(-0.4\,{\rm AB})~.
\end{equation}
Therefore, we transformed the IPHAS magnitudes to the AB scale:
\begin{equation}
r\,({\rm AB}) = r\,({\rm IPHAS}) + 0.163,
\end{equation}
\begin{equation}
i\,({\rm AB}) = i\,({\rm IPHAS}) + 0.387.
\end{equation}
After that the magnitudes in the AB scale were transformed to
$\log \lambda F_{\lambda}$ by the equation:
\begin{equation}
\log \lambda F_{\lambda} = \log
[1.09\,\times\,10^{-5}\,\lambda^{-1}\,\times\,{\rm dex}\,(-0.4\,{\rm
AB})].
\end{equation}

For the {\it Spitzer} passbands the following
equation was used:
\begin{equation}
\log \lambda F_{\lambda} = \log\,(3 \times 10^{-9} \times \lambda^{-1}
\times F_{\nu}),
\end{equation}
with $\lambda$ in $\mu$m and $F_{\nu}$ in Jy. The fluxes in Jy were
obtained from the {\it Spitzer} magnitudes by the equation:
\begin{equation}
m = -2.5 \log\,(F_{\nu} / F_{\nu}^{m=0}),
\end{equation}
where $F_{\nu}^{m=0}$ is the flux of Vega in the passband in Janskys.
The following fluxes of Vega in the passbands $I_1$, $I_2$, $I_3$, $I_4$
and $M_1$ were used: 280.9, 179.7, 115.0, 64.1 and 7.15 Jy.

The spectral energy distributions for 16 stars, for which fluxes in the
MSX and/or {\it Spitzer} systems were available, are shown in Figure 4.
The SEDs for the emission-line stars V521 Cyg, LkH$\alpha$\,189 and
LkH$\alpha$\,191 are shown for comparison.

\sectionb{5}{THE SEARCH FOR X-RAY SOURCES}

Both classical T Tauri stars (CTTS) and post-T Tauri stars with weak
emission lines (WTTS) are emitters of strong X-rays from their active
coronae.  Consequently, the presence of X-ray radiation is one of the
main criteria for the identification of PMS objects.  Therefore we
decided to verify if our stars are associated with any X-ray sources.

According to an X-ray catalog search of the Goddard Space Flight
Center (http://heasarc.gsfc.nasa.gov), X-ray sources in the region
of the NAP nebulae are investigated only by the ROSAT satellite.  ROSAT
has observed all the sky in the 0.1--2.4 keV energy range in a scanning
mode, about 14\% of the sky in a pointed mode and about 1.8\% of the sky
in a pointed high-resolution mode.  The accuracy of the coordinates is
about 1\arcmin\ in the scanning mode, about 10\arcsec\ in the pointed
mode and 1--5\arcsec\ in the pointed high-resolution mode.  The limiting
count rates are 0.01--0.05 ct/s in the scanning mode, 0.001--0.002 ct/s
in the pointed and pointed high-resolution modes with exposures of 1--4
hours.

ROSAT scanning observations cover the whole NAP area but pointed
observations are missing in the SE corner with RA\,$>$20:58 and
DEC\,$<$+44\degr.  High resolution observations are available only in a
circle with a radius of 20\arcmin\ at RA = 20:52, DEC = +44:20
and they cover the Pelican Nebula.

The following ROSAT catalogs available at CDS were searched:  the bright
source catalog (BSC, Voges et al. 1999), the faint source catalog (FSC,
Voges et al. 2000), the catalog of HRI pointed observations (1RXH, ROSAT
Team 2000a) and the 2nd catalog of pointed observations (2RXP, ROSAT
Team 2000b).  It was strange enough to realize that none of our objects
from Table 1 could be found in these catalogs.  But what is more, in the
ROSAT catalogs we could not identify any known NAP T Tauri star either.
In the above mentioned area in the Pelican Nebula, covered by the ROSAT
high-resolution pointing observations, a number of T Tauri and H$\alpha$
emission line stars are known.  None of them were found in the 1RXH
catalog, despite the fact that the limiting photon flux in the area was
about 0.01 ct/s with about 1.3 hour exposure.

Trying to find the explanation, we have checked X-ray counts given by
PMS objects in the Orion association which is roughly at the same
distance from the Sun as the NAP complex.  In Orion, Gagne et al.
(1995) were able to measure with the HRI instrument PMS objects of
spectral types K and M with count rates between 0.001 and 0.03 ct/s
and exposures from 3 to 8 hours.  The limiting count rate 0.001 ct/s
with an 8 hour exposure is equivalent to about 0.006 ct/s with an 1.3
hour exposure used in Pelican.  The brightest sources in Orion (0.03
ct/s)

%%%%%%%%%%%%%%%%%%%% FIGURE 4. SEDS IN 0.55-70 MKM

\vbox{
\centerline{\psfig{figure=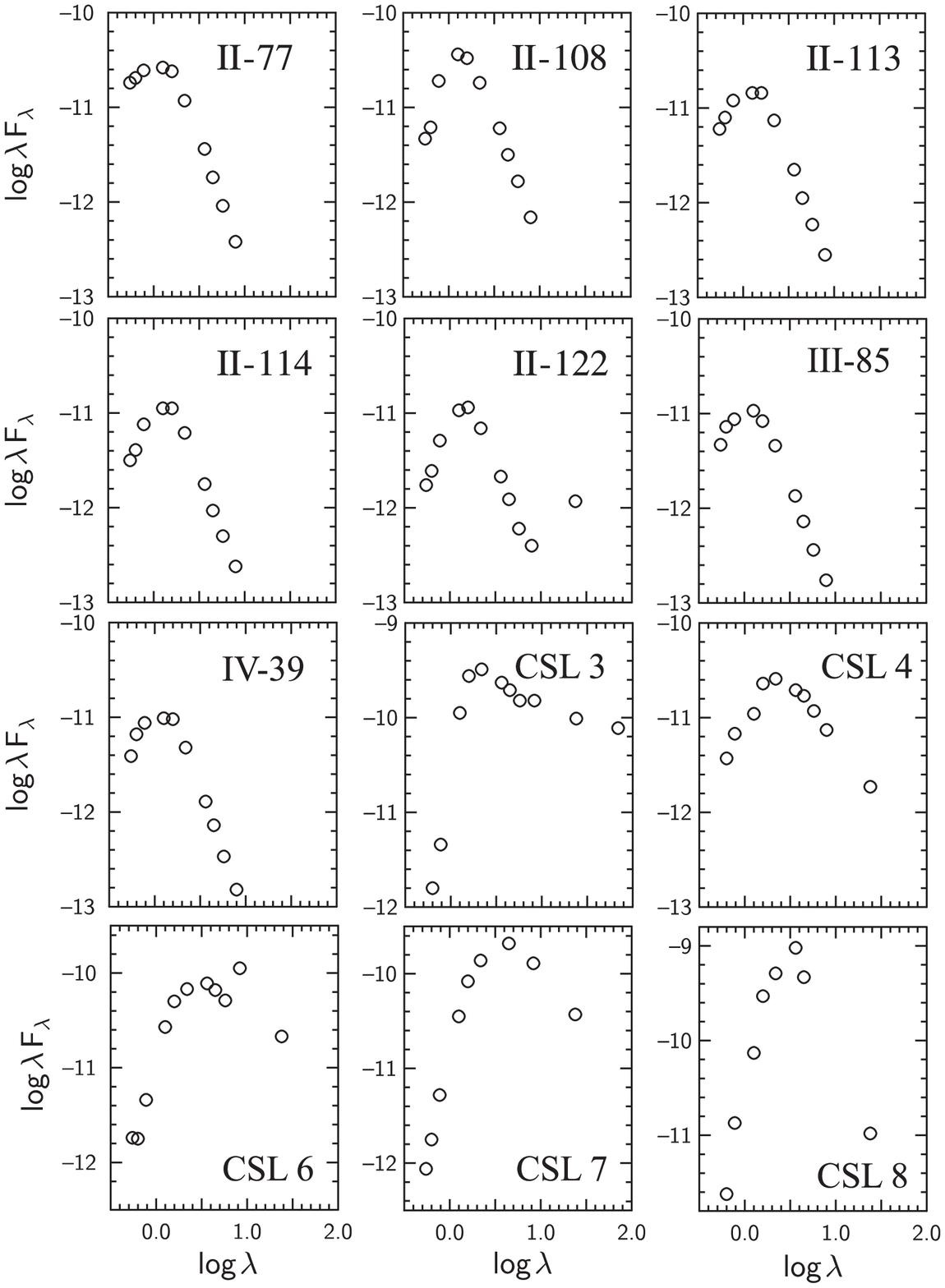,width=124mm,angle=0,clip=true}}
\vskip2mm
\captionc{4}{Spectral energy distributions between 0.5 $\mu$m and 70
$\mu$m.}
\vspace{10mm}
}

\vbox{
\centerline{\psfig{figure=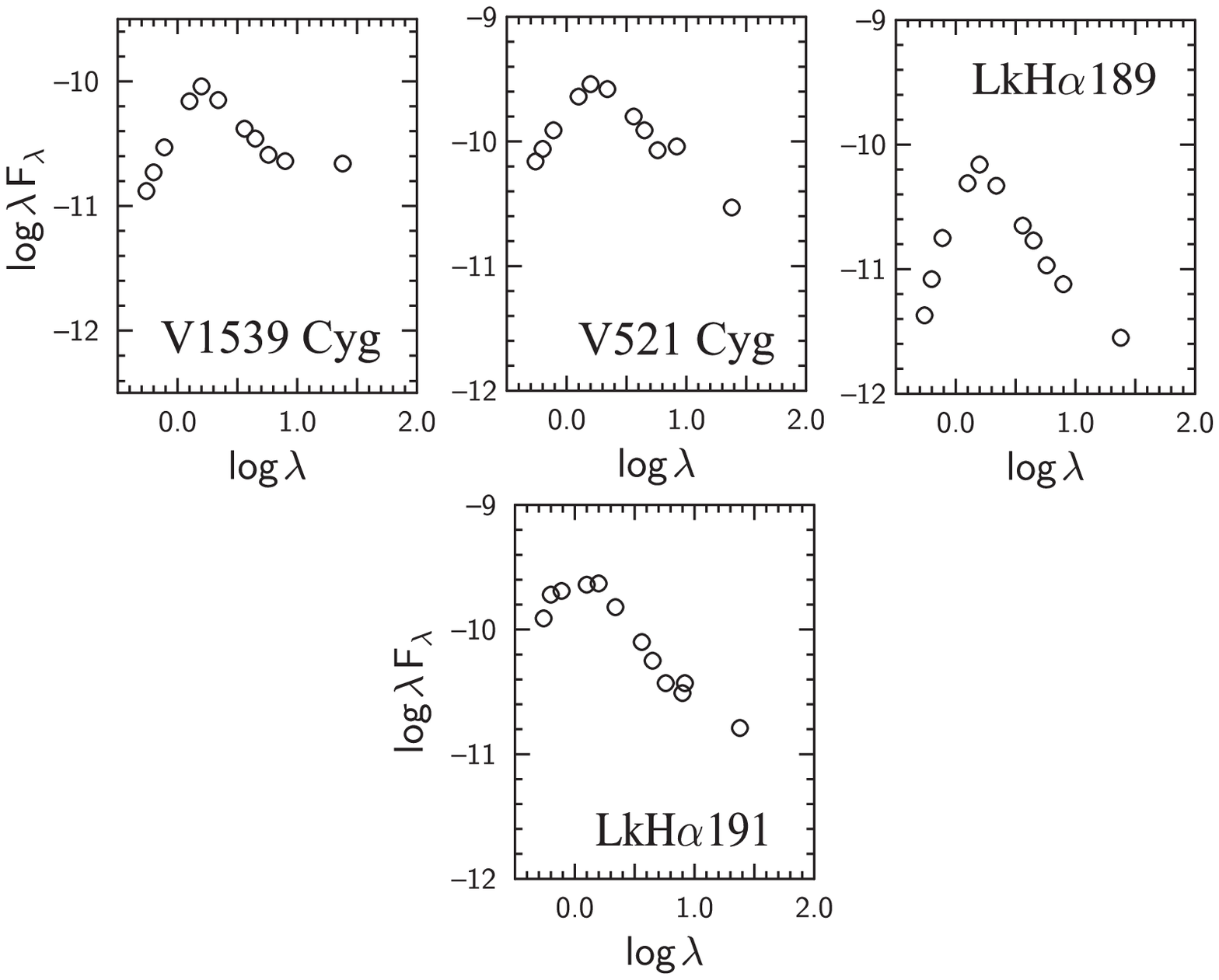,width=124mm,angle=0,clip=true}}
\vspace{-3mm}
\captionc{4}{Continued.}
\vspace{1mm}
}

\noindent
with the Pelican exposure should give 0.005 ct/s which is about
the same as the limiting count rate in Pelican.  Thus, the non-detection
of the Pelican PMS stars in X-rays can be explained by too short HRI
exposures.

At the same time, X-ray fluxes of PMS stars are easily detectable in the
nearby star-forming regions (Tau/Aur, Oph, Lup, Cha) where the X-ray
flux is about 10 times stronger due to their small distances.

%%%%%%%%%%%%%%%%% FIGURES 5 and 6. J-H,H-K DIAGRAMS

\begin{figure}[!th]
\centerline{\psfig{figure=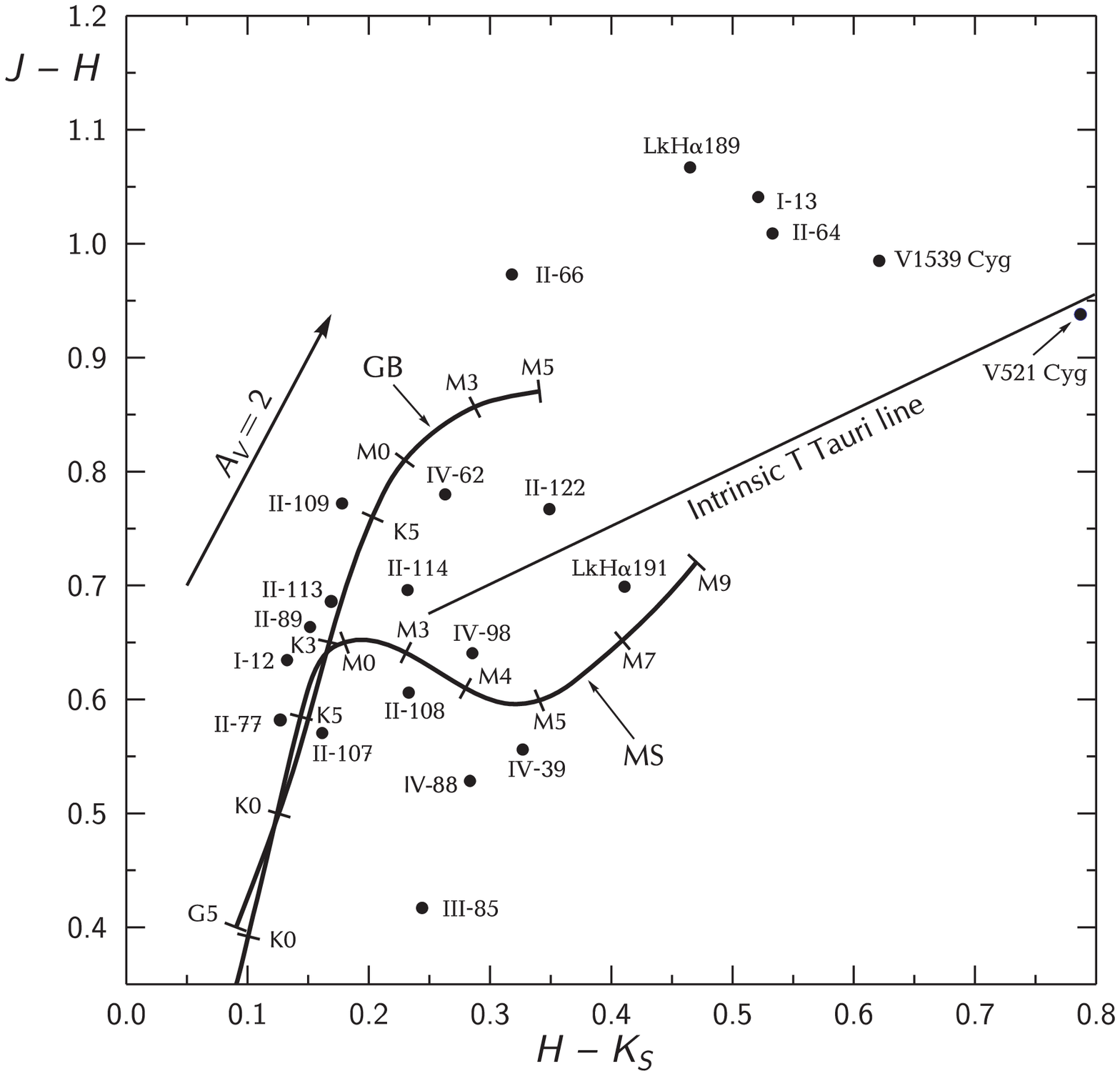,width=124mm,angle=0,clip=true}}
\vspace{0.5mm}
\captionb{5}{$J$--$H$ vs. $H$--$K_s$ diagram for the stars with
H$\alpha$ emission found in their spectra. A few non-emission stars
discussed in the text are also plotted. The main-sequence
(MS), the giant branch (GB) and the T Tauri star intrinsic lines are
shown. The arrow indicates the interstellar reddening vector for $A_V$ =
2 mag.}
\vspace{-1mm}
\end{figure}

\begin{figure}[!th]
\hbox{\parbox[c]{70mm}{\psfig{figure=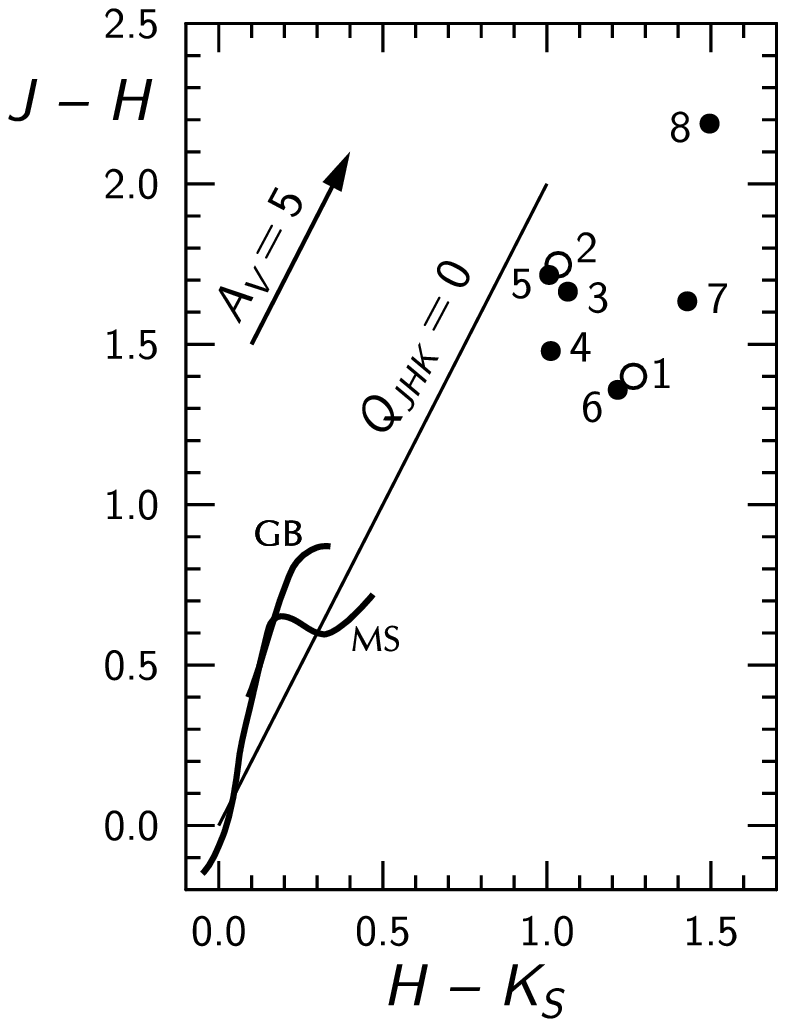,width=70mm,angle=0,clip=true}}
\hskip4mm
\parbox[c]{50mm}{
\captionb{6}{$J$--$H$ vs. $H$--$K_s$ diagram for the investigated stars
with $H$--$K_s$\,$>$\,1.0. The numbers at the points refer to the CSL
numbers in Table 3. The stars with H$\alpha$
emission are plotted as dots, the two non-emission stars as circles. The
intrinsic lines of the main sequence and giants, the interstellar
reddening vector for $A_V$ = 5 mag and the reddening line separating
stars with positive and negative $Q_{JHK}$ values are also shown.}}}
\end{figure}

\sectionb{6}{DISCUSSION OF INDIVIDUAL OBJECTS}

In this section we give more information about the investigated objects
and discuss the results of spectral classification and emission line
intensities.  To understand better photometric properties of the stars
described, in Figures 5 and 6 we show the $J$--$H$ vs. $H$--$K_s$
diagrams for the two groups of the investigated stars.

\vskip1mm

{\bf II-64 = 2MASS J20570757+4341597}

This star, as well as other stars with the numbers starting with II and
III, are located in the Gulf of Mexico of the dark cloud LDN\,935
(TGU\,497).  Based on the interstellar reddening-free $Q_{XZS}$
vs. $Q_{XYV}$ and $Q_{XZS}$ vs. $Q_{YZV}$ diagrams of the {\it
Vilnius} photometric system the star was suspected to be in the PMS
stage of evolution (Laugalys et al. 2006).  In both diagrams the star
deviates considerably from the main sequence, and such its position can
be explained by the presence of emission in the H$\alpha$ line which
makes the star brighter in the $S$ passband centered on this line.  In
the $J$--$H$ vs.\,$H$--$K_s$ diagram (Figure 5) the star lies 0.2 mag
above the intrinsic line of T Tauri stars (Meyer et al. 1997).  Our
spectrum shows that the star is of spectral class K2 with the moderate
emission in H$\alpha$ ({\it EW} = --\,6.8 \AA).  Since the star has no
measurements in the MSX and {\it Spitzer} bands, we have no possibility
to verify the presence of infrared emission at $\lambda$\,$>$\,3 $\mu$m.

The star is only 6\arcsec\ from a slightly brighter star, II-66 = 2MASS
J20570786+\\4341554, classified from {\it Vilnius} photometry as a K5
dwarf.  Its 2MASS colors $J$--$H$ = 0.973 and $H$--$K_s$ = 0.318 place
II-66 $\sim$\,0.25 mag above the intrinsic T Tau line.  Consequently,
the star can be an YSO, too.  This is also confirmed by {\it Spitzer}
photometry (Table 4 of Guieu et al. 2009).

\vskip1mm

{\bf II-77 = 2MASS J20572224+4357534}

In the {\it Vilnius} interstellar reddening-free diagrams the star
considerably deviates from the main sequence, not less than II-64.  In
its spectrum, however, H$\alpha$ emission is much fainter and the
spectral class is earlier (G8).  In the $J$--$H$ vs.  $H$--$K_s$ diagram
(Figure 5) the star lies close to the K dwarf sequence.  The star was
measured in the four {\it Spitzer} IRAC bands, and its energy
distribution (Figure 4) does not show any infrared emission.
Consequently, the star seems to be be a slightly reddened G dwarf with
chromospheric activity.  However, in this case it is difficult to
explain its anomalous position in the diagrams of the {\it Vilnius}
system.  If the star is G8\,V, its distance is 580 pc, i.e., close to
the distance of the NAP complex.

\vskip1mm

{\bf II-108 = 2MASS J20574880+4350236}

In the $Q,Q$ diagrams of the {\it Vilnius} system the star deviates from
the K-dwarf sequence as much as the stars II-64 and II-77.  However, our
spectrum shows that the star differs from the two previous stars being
of spectral class M3.5e and having fainter H$\alpha$ emission ({\it EW}
= --0.2 \AA).  In the $J$--$H$ vs.  $H$--$K_s$ diagram (Figure 5) the
star lies close to the intrinsic position of M3\,V stars with no
infrared excess.  Its energy distribution (Figure 4), constructed from
the 2MASS and {\it Spitzer} observations, also evidences a normal star
without envelope.  Most probably the star is a M-dwarf with weak
chromospheric activity.  Its $V$ = 16.59 and $M_V$ = +10.9 (for a
M3.5\,V star) correspond to a distance from the Sun of 137 pc, i.e., the
star cannot have any relation to the NAP complex.

\vskip1mm

{\bf II-109 = 2MASS J20575005+4350569}

Interstellar reddening-free parameters in the {\it Vilnius} system
classify the star as a dwarf of spectral class M1, but the
classification is of low accuracy since the star is quite faint ($V$ =
17.88) and its ultraviolet magnitudes $U$ and $P$ have not been
measured.  The $S$ magnitude on H$\alpha$ shows a mild indication of
emission. 2MASS color indices of this star, $J$--$H$ = 0.772 and
$H$--$K_s$ = 0.178, place it well above the intrinsic T Tauri star line
(Figure 5).  However, the magnitudes {\it J, H} and $K_s$ are of low
accuracy, with the 0.05--0.07 mag errors.  The star is absent both in
the MSX and in {\it Spitzer} lists, so we cannot verify if the envelope
is present.  In our spectrum it looks like a G-type star with H$\alpha$
filled with emission but this does not help to understand the behavior
of its photometric indices.

\vskip1mm

{\bf II-113 = 2MASS J20575651+4352362}

In the {\it Vilnius} system the star is classified as K dwarf with
possible H$\alpha$ emission.  In the $Q,Q$ diagrams, indicating the
emission, the star lies close to II-109.  Both stars are also similar in
our spectra, they are classified as Ge stars with H$\alpha$ filled with
emission.  In the $J$--$H$ vs.\,$H$--$K_s$ diagram (Figure 5) the star
lies close to the intrinsic position of M0\,V star, but it can be a
reddened G-type star, too.  IRAC magnitudes measured by {\it Spitzer}
show no infrared excess (Figure 4).  Thus, both stars (II-109 and
II-113) could be reddened G main sequence stars but this does not
explain their considerable excesses of radiation in the {\it Vilnius}
$S$ passband. These stars may have been in a particularly active phase
when observed with {\it Vilnius} photometry, and quiet when observed
spectroscopically.

\vskip1mm

{\bf II-114 = 2MASS J20575750+4350089}

In the {\it Vilnius} system the star is classified as M2e. In the
$J$--$H$ vs.\,$H$--$K_s$ diagram (Figure 5) it lies in the upper part of
the early M dwarf band, so there are no indications of the presence of
the envelope.  This is confirmed by the {\it Spitzer} fluxes (Figure 4).
Our spectrum shows that the star is of spectral class M1 with small
emission in H$\alpha$ ({\it EW} = --1.8 \AA).  The star can be either an
unreddened field dwarf with chromospheric activity, located at 350 pc,
or a post-T Tauri star at the front edge of the NAP complex.

\vskip1mm

{\bf II-117 = V\,1539 Cyg = 2MASS J20575986+4353260}

This star is a known emission-line star (LkH$\alpha$\,185), T Tauri type
variable.  The {\it Vilnius} reddening-free parameter $Q_{XZS}$ shows
very strong anomaly caused by strong emission in H$\alpha$, which from
our spectrum gives {\it EW} = --79.4 \AA.  Ca\,II and O\,I lines are
also in emission.  In the $J$--$H$ vs.\,$H$--$K_s$ diagram (Figure 5)
the star lies by 0.12 mag above the intrinsic T Tauri line.  The {\it
Spitzer} fluxes exhibit the presence of a considerable thermal emission
from the dust envelope (YSO of class II, Figure 4).  Our spectrum shows
its spectral class G5e.

\vskip1mm

{\bf II-122 = 2MASS J20580604+4349328}

In the {\it Vilnius} system the star is classified as a K--M dwarf
with possible emission, but due to its faintness ($V$ = 17.66)
the ultraviolet magnitudes $U$ and $P$ were not measured, the
violet $X$ and the blue $Y$ magnitudes are also of low accuracy.  In
the $J$--$H$ vs.\,$H$--$K_s$ diagram (Figure 5) the star lies 0.05 mag
above the intrinsic T Tauri line.  The {\it Spitzer} fluxes exhibit the
presence of emission in {\it I4} and {\it M1} bands (at
$\lambda$\,$\geq$\,8 $\mu$m) confirming the YSO status of the star.
Our spectrum gives the spectral class K7e with strong H$\alpha$ line
({\it EW} = --41.1 \AA). If the star is close to the main sequence, its
distance is similar to the NAP complex.

\vskip1mm

{\bf III-85 = 2MASS J20593535+4352035}

This object was found peculiar by Laugalys et al.  (2006) due to an
excess of radiation in the passband $S$.  In the $J$--$H$
vs.\,$H$--$K_s$ diagram (Figure 5) the star appears to be of early type
with a considerable reddening.  Our spectral classification gives the
spectral class A0 with the emission component inside the H$\alpha$ line,
shifted longward from the absorption, or the strong absorption shifted
shortward of emission (the P Cygni profile).  The SED curve in the
infrared does not show any excess (Figure 4).  The star III-85 is far
behind the NAP complex, see the description of the following star.

\vskip1mm

{\bf IV-5 = 2MASS J20535187+4424100}

This and the following three stars are located in Area IV (Laugalys et
al. 2006), above Pelican's beak, near the two bright stars, HD\,199081
(57 Cyg) and HD\,199178.  The star IV-5 could not be classified in the
{\it Vilnius} system -- its color indices are very peculiar.  A
considerably diminished $Q_{XZS}$ parameter was in favor of the presence
of H$\alpha$ emission. 2MASS magnitudes are given with a very low
accuracy and are useless.  No doubt, its image was affected by a nearby
extremely red AGB star 2MASS J20535282+4424015 = StRS\,383 discussed by
Neckel et al.  (1980) and Eiroa et al.  (1983) and included into a list
of highly reddened stars by Stephenson (1992).  The brightness of the
latter star drastically increases with increasing wavelength:  its blue
magnitude is about 20, $V$ = 16.7, $N$\,(840 nm) = 10.4, $J$ = 6.2, $H$
= 4.3, $K$ = 3.2 and $L$ = 2.0.  Thus, in the 2MASS survey the star IV-5
is immersed in the oversaturated image of StRS\,383.  We classified the
spectrum of IV-5 as A0e, it is quite similar to that of the star III-85
with strong H$\alpha$ absorption line having a redshifted emission
component.  Both these stars should be farther than 2.7 kpc.

\vskip1mm

{\bf IV-39 = 2MASS J20542554+4423019}

In the {\it Vilnius} system the star is classified as a K or M dwarf
with emission.  In the $J$--$H$ vs.\,$H$--$K_s$ diagram (Figure 5) the
star lies on the sequence of M-dwarfs.  Our spectral classification
gives the spectral class M3 with a faint emission in H$\alpha$ ({\it EW}
= --1.2 \AA).  The SED curve in the infrared does not show any excess
(Figure 4).  Consequently, the star seems to be a chromospherically
active M dwarf.  For $V$ = 17.37, $M_V$ = +10.4 and no reddening we
obtain a distance of 250 pc, i.e., it is much closer than the NAP
complex.

\vskip1mm

{\bf IV-62 = 2MASS J20544534+4433027}

In the {\it Vilnius} system the star is classified as a reddened K0
dwarf with emission.  In the $J$--$H$ vs.\,$H$--$K_s$ diagram (Figure 5)
the star lies 0.1 mag above the intrinsic T Tauri line.  Our spectral
classification gives the spectral class K0 with faint emission in
H$\alpha$ ({\it EW} = --1.8 \AA).  With $V$ = 17.24, $M_V$ = +5.9 and
$A_V$ = 3.21 (Laugalys et al. 2006) its distance is 420 pc, i.e.
close to the distance of the NAP complex.  Thus, the star can
be either a chromospherically active field dwarf or a post T Tauri star
immersed in the dark cloud.

\vskip1mm

{\bf IV-98 = 2MASS J20551122+4424052}
In the {\it Vilnius} system the star is classified as B star with
emission. In the $J$--$H$ vs.\,$H$--$K_s$ diagram (Figure 5) the star
lies close to the M4\,V position where the star could appear due to
large interstellar reddening. Our spectrum around H$\alpha$ shows that
it could be a mid-B star, if H$\alpha$ is partly filled-in by emission.
Probably the star is a mild Be star with $A_V$\,$\approx$\,6.7. If we
accept it is a B5 star on the main sequence, its distance is about
1.2 kpc.

\vskip1mm

{\bf CSL\,3 = 2MASS J20485070+4349496}

The star is located in the direction of the Pelican nebula.  Its colors
$J$--$H$ = 1.66, $H$--$K_s$ = 1.06 place the star $\sim$\,0.5 mag above
the intrinsic T Tauri line, with $Q_{JHK_s}$ = --0.46 (Figure 6).  The
IRAS 20472+4338 source with the coordinates corresponding to an angular
distance of 101\arcsec\ from CSL\,3 probably is not related to this
star.  Our spectrum shows strong emission lines of H$\alpha$ ({\it EW} =
--30.5 \AA) and Ca\,II, the spectral class is G0e.  Spectral energy
distribution (Figure~4) is typical of YSO of Class II (T Tauri
star).  If the star is located in the NAP complex (500 pc) and has
$A_K$\,$\approx$\,2 mag, its absolute magnitude $M_K$ should be close to
--2 mag.

\vskip1mm

{\bf CSL\,4 = 2MASS J20493219+4417031}

The star is located also in the direction of Pelican.  The colors
$J$--$H$ = 1.48 and $H$--$K_s$ = 1.01 place the star $\sim$\,0.4 mag
above the intrinsic T Tauri line, with $Q_{JHK_s}$ = --0.54 (Figure 6).
Our spectrum shows extremely strong emission in H$\alpha$ ({\it EW} =
--151.3 \AA) and strong emissions in Ca\,II, the spectral class is F0e.
The spectral energy distribution (Figure 4) shows an increase of
intensity in the mid-infrared up to the {\it Spitzer} MIPS passband at
24 $\mu$m, but this increase is not so striking as in CSL\,3.  This
seems to be in contradiction with its strong emission lines in the far
red spectrum.  Based on {\it Spitzer} photometry the star was also
included in the list of potential YSOs (Guieu et al. 2009).  At a
distance of 500 pc and $A_K$\,$\approx$\,2 mag, its absolute magnitude
$M_K$ should be close to +0.9 mag.

\vskip1mm

{\bf CSL\,5 = 2MASS J20533846+4428484}

The star is located in the direction of Pelican, close to the star IV-5
described above.  The colors $J$--$H$ = 1.72 and $H$--$K_s$ = 1.01 place
the star $\sim$\,0.6 mag above the intrinsic T Tauri line, its
$Q_{JHK_s}$ = --0.30 (Figure 6).  However, this position is the result
of considerable moving along the reddening line starting from A-type
stars.  Our classification gives the spectral class A5 with the emission
component inside the H$\alpha$ line, longward from the absorption (the P
Cygni like profile).  In this respect the star is quite similar to
III-85 and IV-5 described above.  Since the star was observed
neither by MSX nor {\it Spitzer}, we have no possibility to construct
its spectral energy distribution in the mid-infrared.  The star can be
in a post Herbig Ae evolutionary stage with a remnant disk or envelope.
At a distance of the NAP complex and $A_K$\,$\approx$\,1.9 mag, its
absolute magnitude $M_K$ should be +0.3 mag.

\vskip1mm

{\bf CSL\,6 = 2MASS J20544690+4448197}

The star is located at the upper right edge of the North America nebula.
Its $J$--$H$ = 1.36 and $H$--$K_s$ = 1.22 place the star close to the
intrinsic T Tauri line, with $Q_{JHK_s}$ = --1.07 (Figure 6).  Our
spectrum shows that this is a heavily reddened Ae star with strong
emission lines of H$\alpha$ ({\it EW} = --82.9 \AA) and Ca\,II.  The
spectral energy distribution (Figure~4) shows an enhanced intensity in
the IRAC and MIPS bands which is typical of YSO of Class III.  The MSX
8.3 $\mu$m band measurement gives much brighter value
($\log\,\lambda\,F_{\lambda}$ = --9.95) than the interpolated {\it
Spitzer} data.  Maybe, this is related with the IR emission variability.
We conclude that CSL\,6 is a typical medium-mass Herbig Ae-type YSO.
The star is recognized as an emission star in the IPHAS survey.  At a
distance of the NAP complex and $A_K$\,$\approx$\,2.4 mag, its $M_K$ is
close to --0.6 mag.

\vskip1mm

{\bf CSL\,7 = 2MASS J20562782+4525234}

The star is located in the direction of the upper part of the North
America nebula.  Its $J$--$H$ = 1.63 and $H$--$K_s$ = 1.43 place the
star above the intrinsic T Tauri line, with $Q_{JHK_s}$ = --1.22 (Figure
6).  However, the star has no relation to T Tauri stars, being a heavily
reddened Herbig Ae star of medium-mass.  Our spectrum shows the
strongest emission of H$\alpha$ among all of the investigated stars
({\it EW} = --168.7 \AA).  Emission lines of O\,I, Ca\,II and P9 are
also strong, spectral class is close to A0.  The spectral energy
distribution (Figure~4) shows very strong interstellar (and
circumstellar) reddening and the presence of enhanced intensity in the
IRAC and MIPS bands, typical of YSOs of Class III.  At a distance of the
NAP complex and $A_K$\,$\approx$\,2.8 mag, its $M_K$ is close to --1.8
mag.

\vskip1mm

{\bf CSL\,8 = 2MASS J21010487+4342340}

The star is located in the direction of a bright nebula to the left from
the Gulf of Mexico.  Its $J$--$H$ = 2.19 and $H$--$K_s$ = 1.50 place the
star about 0.75 mag above the intrinsic T Tauri line, with $Q_{JHK_s}$ =
--0.80 (Figure 6).  Our spectral classification shows that it is a
K3 supergiant with moderate emission in H$\alpha$ ({\it EW} = --5.0
\AA).  Its spectral energy distribution (Figure~4) shows a very large
interstellar reddening, with maximum in the 3.6 $\mu$m IRAC band.
Very strong interstellar reddening is essential to explain its position
in the $J$--$H$ vs.\,$H$--$K_s$ diagram.  If the star is K3\,Iab, its
$M_K$\,$\approx$\,--9.8 mag.  With $K_s$ = 8.11 and $A_{K_s}$ = 2.7 mag
we get a distance of 11 kpc, i.e., the star can be located in the
Outer spiral arm.  The H$\alpha$ emission seen in the spectrum can
originate in a nebula at the close vicinity of the supergiant.

\vskip1mm

{\bf Stars without emission lines}

In the spectra of some stars, which in the paper Laugalys et al.
(2006) were suspected to have H$\alpha$ in emission,  emission was not
found.  These stars are listed in Table 3, where the fifth column gives
spectral classes estimated from our spectra.  Most of these stars were
found to be M-dwarfs, not K-dwarfs as they were classified in the {\it
Vilnius} system.  This may mean that K and M dwarfs are not sufficiently
represented among the standard stars used to classify stars from the
photometric data.  On the other hand, most of these stars are fainter
than $V$ = 16 mag, and for many of them ultraviolet magnitudes $U$ and
$P$ were too faint to be measured with good accuracy.

It should be noted that the classification of some of these stars
contradicts their positions in the $J$--$H$ vs.\,$H$--$K_s$ diagram.
For example, the star I-13 lies about 0.2 mag above the intrinsic T
Tauri line.  If the star is a M2 dwarf, it should be considerably
shifted along the interstellar reddening line ($A_V$\,$\approx$\,4 mag).
However, the spectrum of I-13 does not show such reddening.  The stars
I-12, II-89 and II-107 have too blue $H$--$K_s$ colors for the spectral
class dM2.  The star IV-88, if it were G dwarf, should have considerable
reddening in agreement with its spectrum shape (unfortunately, quite
noisy).  The infrared colors of CSL\,1 and CSL\,2 also do not agree with
the finding that these stars are K0 and G8 dwarfs:  none of normal G--K
stars can have such negative values of $Q_{JHK}$ (Figure 6).  In some
cases these discrepancies can be explained by unresolved duplicity of
the stars.  For example, the image of the star I-12 in DSS2 is clearly
asymmetric, this suggests the star to be a binary.  The stars CSL\,1 and
CSL\,2 can be physical binaries with L-dwarf components which might be
responsible for their strong infrared excesses.

Assuming that eleven K and M stars in Table 3 are of luminosity V, we
get their interstellar extinctions and distances given in the last two
columns.  It is evident that all these stars are located in front of the
LDN\,935 cloud.  The luminosity class of the III-71 star of Table 3 is
not known but it is evident that it is a distant object located far
behind the NAP complex.

%%%%%%%%%%%%%%%%%%%%%% TABLE 3. STARS WITHOUT EMISSION
\begin{table}[!t]
\begin{center}
\vbox{\small\tabcolsep=6pt
\parbox[c]{100mm}{\baselineskip=10pt
{\normbf\ \ Table 3.}{\norm\ Stars for which our spectra
do not show the presence of strong emission lines.
\lstrut}}
\begin{tabular}{lcccccc}
\tablerule
 Name  &  RA\,(2000) & DEC\,(2000) &  $V$ & Sp & $A_V$ & $d$~(pc) \\
\tablerule
I-12*   & 20:54:40.64 & +43:52:47.2 & 17.2 & dM2.5 & 0.62 & 184 \\
I-13*   & 20:54:42.08 & +43:45:57.3 & 17.5 & dM2   & 0.46 & 268 \\
I-53   & 20:55:27.48 & +43:53:17.8 & 17.5 &  dM1   & 1.08 & 260 \\
I-61   & 20:55:33.59 & +43:52:19.8 & 16.8 &  dM1   & 0.62 & 233 \\
I-62   & 20:55:34.93 & +43:49:44.2 & 16.0 &  dM3.5 & 0.21 & ~~97 \\
II-50  & 20:56:51.69 & +43:42:52.0 & 15.3 &  dK3   & 0.71 & 400 \\
II-67  & 20:57:09.06 & +43:44:44.5 & 17.3 &  dM3   & 0.00 & 240 \\
II-89*  & 20:57:28.64 & +43:58:31.6 & 16.3 & dM2   & 0.46 & 156 \\
II-102 & 20:57:43.26 & +43:56:31.9 & 16.6 &  dM4   & 0.00 & 108 \\
II-107* & 20:57:48.00 & +43:42:54.6 & 17.4 & dM2   & 0.50 & 256 \\
II-118 & 20:57:59.93 & +43:51:21.2 & 16.8 &  dM2   & 0.62 & 178 \\
III-71 & 20:59:29.29 & +43:45:56.5 & 16.8 &  A0    &      &     \\
IV-88** & 20:55:05.18 & +44:36:31.1 & 17.9 & G8:   &      &     \\
CSL\,1* & 20:46:46.94 & +43:17:34.6 & 17.5 & dK0 &      &     \\
CSL\,2* & 20:48:30.50 & +44:22:54.6 & 17.0(r) & dG8 &    &     \\
\tablerule
\end{tabular}
}
\end{center}
\parbox[c]{120mm}{\baselineskip=7pt\footnotesize
\noindent {\bf Notes.}~~* These stars are
discussed in the text; ** IV-88, noisy spectrum.}
\vskip-.2mm
\end{table}

\sectionb{7}{CONCLUSIONS}

The present spectral investigation shows that the suspected H$\alpha$
emission stars in the NAP complex do not constitute a homogeneous group.
Fourteen of them were found to be normal G-, K- and M-dwarfs without
emission.  This may indicate that their emission in H$\alpha$ is
variable and sometimes disappears.  An alternative possibility is
related to the errors of photometric classification due to insufficient
accuracy of CCD photometry or calibration of photometric parameters of
red dwarfs (or both).  In some cases, the peculiarity of photometric
color indices can be the result of unresolved binarity.  Two stars
without emission, CSL\,1 and CSL\,2, are very strong in the near
infrared; their infrared excesses are far too large to be of
interstellar origin.  These stars can be binaries with the cool
components of late M or L spectral types.  The infrared excess of I-13
may be of the same origin.

The remaining 19 stars in our list exhibit different strengths of
H$\alpha$ emission and different spectral types.  Among these stars
seven show strong emission, six show medium and weak emission, and in
six spectra the H$\alpha$ line is filled by emission or has an emission
component.  In spectral classes we have the following distribution:  one
is a mid-B star, five A stars, one F star, five G stars, four K stars
and three M stars.  For six stars the mid-infrared thermal emission in
the dust envelope was found; seven stars do not show the presence of
such emission and for six stars the data were missing.

Only eight stars in the investigated emission-line star sample seem to
be members of the NAP complex.  Five of them belong to the group of
eight suspected YSOs with negative values of $Q_{JHK_s}$ and
$H$--$K_s$\,$>$\,1.0.  This means that a considerable fraction of
true YSOs, related to the NAP complex, is expected among stars with
negative $Q_{JHK_s}$ values.  Tens of such objects in our list selected
from 2MASS are fainter than magnitude 17 and they were not accessible to
the telescope used.  Many more YSOs should be present among the objects
with $H$--$K_s$ between 0.5 and 1.0.  Consequently, the prediction of a
large PMS population in the NAP complex still holds to be true.
A big list of possible YSOs in the region, suspected with the help of
{\it Spitzer} photometry (Guieu et al. 2009), is a confirmation.

The most effective methods for their detection are deep surveys of
H$\alpha$ emission stars using photometry in the {\it Vilnius} or IPHAS
systems.  To confirm the evolutionary status of the suspected young
objects we need their red spectra for measuring H$\alpha$ intensity and
also for checking the presence of the Li\,I doublet at 6708 \AA\ which
is a nice test of the star youth.  Although in our spectra of some stars
the Li line is visible, for its exact identification and {\it EW}
estimation a better {\it S/N} ratio is desirable.  Mid-infrared
photometry in $\lambda$\,$>$\,5 $\mu$m is essential to verify the
presence of circumstellar envelopes, and X-ray observations can confirm
the presence of active coronae, a characteristic feature among T~Tauri
and post-T Tauri stars.

\thanks{ We are thankful to the Steward Observatory for the observing
time, to Luisa Rebull for the {\it Spitzer} data and to Edmundas
Mei\v{s}tas and Stanislava Barta\v{s}i\={u}t\.e for their help in
preparing the paper.  The use of the 2MASS, MSX, IPHAS, SkyView, Gator
and Simbad databases and the IRAF program package is acknowledged.}

\References

\refb Comer\'{o}n F., Pasquali A. 2005, A\&A, 430, 541

\refb Corbally C. J., Strai\v{z}ys V. 2008, Baltic Astronomy, 17, 21

\refb Corbally C. J., Strai\v{z}ys V. 2009, Baltic Astronomy, 18, 1

\refb Danks A. C., Dennefeld M. 1994, PASP, 106, 382

\refb Drew J. E., Greimel R., Irwin M. J. et al. 2005, MNRAS, 362, 753

\refb Eiroa C., Hefele H., Zhong-yu Q. 1983, A\&AS, 54, 309

\refb Egan M. P., Price S. D., Kraemer K. E. et al. 2003, {\it The
Midcourse Space Experiment Point Source Catalog}, version 2.3,
AFRL-VS-TR-2003-1589; available at CDS: MSX6C Infrared Point Source
Catalog, V/144

\refb Gagn\'e M., Caillaut J.-P., Stauffer J. R. 1995, ApJ, 445, 280

\refb Gieseking F. 1973, Ver\"off. Astron. Inst. Bonn, Nr. 87

\refb Gieseking P., Schumann J. D. 1976, A\&AS, 26, 367

\refb Gonz\'alez-Solares E. A., Walton N. A., Greimel R., Drew J. E.
2008, MNRAS, 388, 89

\refb Guieu S., Rebull L. M., Stauffer J. R. et al. 2009, ApJ, 697, 767

\refb Herbig G. H. 1958, ApJ, 128, 259

\refb Kohoutek L., Wehmeyer R. 1997, {\it H-alpha Stars in Northern
Milky Way}, Abh. Hamburger Sternw., 11, Teil 1+2 = CDS catalog III/205

\refb Lasker B. M., Lattanzi M. G., McLean B. J. et al. 2008, AJ, 136,
735; Simbad Catalog I/305, version GSC\,2.3.2

\refb Laugalys V., Strai\v{z}ys V., Vrba F. J., Boyle R. P., Philip
A.\,G\,D., Kazlauskas A. 2006, Baltic Astronomy, 15, 483

\refb Marcy G. W. 1980, AJ, 85, 230

\refb Meyer M. R., Calvert N., Hillenbrand L. A. 1997, AJ, 114, 288

\refb Neckel Th., Harris A. W., Eiroa C. 1980, A\&A, 92, L9

\refb ROSAT Team 2000a, ROSAT News, No.71;
available at CDS: The ROSAT Source Catalog of Pointed
Observations with the High Resolution Imager (HRI, 1RXH), IX/28A

\refb ROSAT Team 2000b, ROSAT News, No.72;
available at CDS: The ROSAT Source Catalog of Pointed
Observations with the Position Sensitive Proportional Counter
(PSPC, 2RXH), IX/30

\refb Stephenson C. B. 1992, AJ, 103, 263

\refb Strai\v{z}ys V., Corbally C. J., Laugalys V. 2008, Baltic
Astronomy, 17, 125

\refb Strai\v{z}ys V., Eimontas A., S\={u}d\v{z}ius et al. 1998, Baltic
Astronomy, 7, 589

\refb Strai\v{z}ys V., Laugalys V. 2007, Baltic Astronomy, 16, 327

\refb Strai\v{z}ys V., Laugalys V. 2008, Baltic Astronomy, 17, 143

\refb Tsvetkov M. K. 1975, Astrofizika, 11, 579

\refb Voges W., Aschenbach B., Boller Th. et al. 1999, A\&A, 349, 389;
available at CDS: ROSAT All-Sky Survey Bright Source Catalog (1RXS),
IX/10

\refb Voges W., Aschenbach B., Boller Th. et al. 2000, IAUC, 7432;
available at CDS: ROSAT All-Sky Survey Faint Source Catalog, IX/29

\refb Welin G. 1973, A\&AS, 9, 183

\refb Witham A. R., Knigge C., Drew J. E. et al. 2008, MNRAS, 384, 1277

\end{document}